\documentclass[10pt,journal,compsoc]{IEEEtran}
\usepackage{graphicx}
\usepackage{url}
\usepackage{verbatim}
\usepackage{float}
\usepackage{subfig}
\usepackage{lineno}
\usepackage{comment}
\usepackage{booktabs}
\usepackage{amsmath,epsfig,psfrag}
\usepackage{array}
\newcolumntype{P}[1]{>{\centering\arraybackslash}p{#1}}

\usepackage[T1]{fontenc}
\usepackage{lmodern}
\usepackage{xcolor}
\usepackage{multirow}
\usepackage{rotating}
\usepackage[normalem]{ulem}

\usepackage{enumitem}
\usepackage{comment}
\usepackage{algorithm,algpseudocode}
\begin{document}
\title{Proactive Defense for Internet-of-Things: Integrating Moving Target Defense with Cyberdeception}

\author{Mengmeng Ge,~\IEEEmembership{Member,~IEEE,}
        Jin-Hee Cho,~\IEEEmembership{Senior Member,~IEEE,}
        Dong Seong Kim,~\IEEEmembership{Member,~IEEE}
        Gaurav Dixit,~\IEEEmembership{Non-Member,~IEEE}
        and
        Ing-Ray Chen~\IEEEmembership{Member,~IEEE}
}
\maketitle
\begin{abstract}

Resource constrained Internet-of-Things (IoT) devices are highly likely to be compromised by attackers because strong security protections may not be suitable to be deployed. This requires an alternative approach to protect vulnerable components in IoT networks. In this paper, we propose an integrated defense technique to achieve intrusion prevention by leveraging cyberdeception (i.e., a decoy system) and moving target defense (i.e., network topology shuffling). We verify the effectiveness and efficiency of our proposed technique analytically based on a graphical security model 
in a software defined networking (SDN)-based IoT network. We develop four strategies (i.e., fixed/random and adaptive/hybrid) to address ``when'' to perform network topology shuffling and three strategies (i.e., genetic algorithm/decoy attack path-based optimization/random) to address ``how'' to perform network topology shuffling on a decoy-populated IoT network, and analyze which strategy can best achieve a system goal such as prolonging the system lifetime, maximizing deception effectiveness, maximizing service availability, or minimizing defense cost. 
Our results demonstrate that a software defined IoT network running our intrusion prevention technique at the optimal parameter setting prolongs system lifetime, increases attack complexity of compromising critical nodes, and maintains superior service availability compared with a counterpart IoT network without running our intrusion prevention technique. Further, when given a single goal or a multi-objective goal (e.g., maximizing the system lifetime and service availability while minimizing the defense cost) as input, the best combination of ``when'' and ``how'' strategies is identified for executing our proposed technique under which the specified goal can be best achieved.
\end{abstract}




\section{Introduction}
Internet-of-Things (IoT) has received significant attention due to their enormous advantages. Advances in IoT technologies can be easily leveraged to maximize effective service provisions to users. However, due to the high heterogeneity and resource constraints of composed entities in a large-scale network, we face the following challenges~\cite{Roman2013ComNet}: (1) distributed technologies for communications, data filtering, processing, and dissemination with 
various forms of data (e.g., text, voice, haptics, image, video) in a large-scale IoT network with heterogeneous entities (i.e., devices, humans); (2) severely restricted resources in battery, computation, communication (e.g., bandwidth), and storage, causing significant challenges in resource allocation and data processing capabilities; (3) highly adversarial environments with compromised, deceptive entities and data, which may result in detrimental impacts on the capabilities of critical mission-related decision making; and (4) highly dynamic interactions between individual entities, data, and environmental factors (e.g., network topology or resource availability), where each factor itself is also highly dynamic in time/space. Due to these characteristics of IoT environments, highly secure, lightweight defense mechanisms are in need to protect and defend the system (or network) against potential attacks.  As a solution to protect and defend a system against inside attacks, many intrusion detection systems (IDSs) have been developed to identify and react to the attacks. However, the core idea of IDSs is reactive in nature and even though it detects intrusions which have already been in the system. Hence, this reactive mechanism normally would be late and ineffective reacting to actions by agile and smart attackers. To overcome the inherent limitation of IDSs due to this reactive nature, intrusion prevention systems (IPSs) have been developed to thwart potential attackers and/or mitigate the impact of the intrusions before they penetrate the system~\cite{cho2018JDMS}. In this work, we are interested in developing an integrated intrusion prevention mechanism based on cyberdeception (i.e., a decoy system) and moving target defense (MTD) and evaluating their effectiveness and efficiency 
by a graphical security model (GSM)-based evaluation framework in a Software Defined Networking (SDN)-based IoT network via simulation. 

\subsection{Research Goal \& Contributions}
This work proposes an integrated proactive defense system based on cyberdeception and moving target defense (MTD) techniques as intrusion preventive mechanisms to minimize the impact of potential attackers trying to penetrate into IoT systems via multiple entries.
This work has the following unique contributions relative to the current state-of-the-art:
\begin{itemize}
\item We are the first to propose an integrated proactive defense system 
by shuffling the topology of an IoT network consisting of both decoy nodes and real nodes. As decoy nodes are part of a decoy system, 
this work integrates MTD with cyberdeception to 
create maximum hurdles and/or complexity to the attackers
while minimizing the defense cost for executing MTD operations. The key goal of the proposed network topology shuffling-based MTD technique (called NTS-MTD henceforward) with decoy nodes is to generate a network topology that can maximize disadvantages against the attackers by misleading attackers to spend time and energy on decoy nodes. There is no prior research in the literature that considers integrating cyberdeception and MTD particularly in the context of network topology shuffling of decoy-populated IoT networks.
\item We address the issues of ``when'' to perform network topology shuffling and ``how'' to perform network topology shuffling on a decoy-populated IoT network. We develop four strategies (fixed/random/adaptive/hybrid) to address ``when'' to perform network topology shuffling and three strategies (genetic algorithm/decoy attack path-based optimization/random) to address ``how'' to perform network topology shuffling and analyze which strategy can best achieve a system goal such as prolonging the system lifetime, maximizing deception effectiveness, maximizing service availability, or minimizing defense cost. Further, when given a single goal (e.g., prolonging the system lifetime) or a multi-objective goal (e.g., maximizing the system lifetime and service availability while minimizing the defense cost) as input, the best combination of ``when'' and ``how'' strategies is identified for executing our proposed technique under which the specified goal can be best achieved.
\item We develop a graphical security model (GSM)  
to evaluate the proposed cyberdeception and MTD technique. The GSM offers design solutions to consider attack graphs (AGs) and/or attack trees (ATs) which can provide efficient methods to calculate the potential security (or vulnerability) levels of attack paths. This  
allows us to analytically evaluate the effectiveness of the proposed cyberdeception and MTD integrated technique in a large IoT network.

\item We consider a Software Defined Networking (SDN)-based IoT system as our network environment. The merits of SDN technology are programmability and controllability, allowing us to develop cyberdeception and MTD integrated technique over a wide range of conditions. We obtain security and performance measures, including the number of attack paths toward decoy targets, mean time to security failure (i.e., MTTSF or system lifetime), and defense cost.

\end{itemize}

A preliminary version of this work appeared in ~\cite{Ge2019Book}. We have substantially extended ~\cite{Ge2019Book} in algorithm design and evaluation, including:
\begin{itemize}
\item We develop a new ``when-to-shuffle'' strategy based on adaptive shuffling and we consider four ``when-to-shuffle'' triggering strategies, viz., fixed, random, adaptive, and hybrid. 
\item We develop a new ``how-to-shuffle'' strategy based on decoy attack path-based optimization for maximizing the number of attack paths toward decoy targets and we consider three ``how-to-shuffle'' strategies, viz., genetic algorithm, decoy attack path-based optimization, and random.
\item We consider a new metric, packet delivery ratio, to measure the service availability in the presence of attacks.
\item We conduct a comparative performance analysis for 12 schemes resulting from a combination of four ``when-to-shuffle'' triggering strategies and three ``how-to-shuffle'' strategies. 
We add a new section to analyze the effects of key parameters, including attack intelligence, attack severity, and decoy/real node population, on system performance. We also add a new section to analyze the sensitivity of performance results with respect to the system security vulnerability level threshold parameter and the maximum delay parameter used in hybrid shuffling.
\end{itemize}

\subsection{Structure of This Paper}
The rest of this paper is organized as follows. Section \ref{sec:related-work} provides a brief overview of the related work in terms of MTD and cyberdeception techniques for IoT environments, security models and metrics, and SDN technology and its use for IoT environments. Section \ref{sec:system-model} gives an overview of the system model, including 
the targeted network environment, node characteristics, attack behaviors, defense mechanisms, and security failure conditions. Section \ref{sec:proactive-defense} describes the design of our proposed integrated proactive defense mechanism in detail, including the strategies of when and how-to-shuffle an IoT network populated with real and decoy nodes to achieve performance goals and the graphical security model used for security analysis.
Section ~\ref{sec:results} shows evaluation results and analyzes the results observed. Section~\ref{sec:conclusions} summarizes key findings and suggests future research directions.

\section{Related Work}\label{sec:related-work}
We briefly survey related work in three areas: (1) existing MTD and cyberdeception techniques for IoT; (2) security models and metrics; and (3) SDN technology for IoT.


\subsection{MTD and Defensive Deception Techniques for IoT} \label{subsec:related-motd-cyberdeception}
The concept of moving target defense (MTD) has been emerged to support the goal of proactive intrusion prevention. The basic idea behind MTD is to defense against attackers by continuously changing attack surface (e.g., system/network configurations) so as to increase attack complexity/cost and also invalidate the system intelligence collected by the attackers~\cite{Cho20-mtd-survey, Hong2015TDSC}.  MTD has been discussed with three main classes: shuffling, diversity, and redundancy. {\em Shuffling-based MTD} aims to confuse attackers by changing network/system configurations such as network addresses (e.g., IP addresses, MAC addresses, or port numbers), software migration, or network topology configuration.  {\em Diversity-based MTD} increases attack complexity by using various types of system components (e.g., software) which provide same functionalities (e.g., using different kinds of operating systems).  Lastly, {\em Redundancy-based MTD} provides security protections by dynamically using multiple replicas of system components in a network for the purpose of maintaining high system reliability~\cite{Cho20-mtd-survey}.  

Several existing MTD techniques have been developed to provide security protection for resource-constrained IoT environments.  \cite{Ge2017JNCA} investigated address space layout randomization (ASLR) and evaluated its performance using the proposed Hierarchical Attack Representation Model (HARM). 
Several lightweight MTD techniques are also proposed by randomly choosing different types of cryptographic primitives~\cite{Plaga2018ICST} or both cryptosystems and firmwares~\cite{Casola2013AMM} for wireless sensor networks.  
\cite{sherburne2014ACISRC} proposed a dynamically changing IPv6 address assignment approach over the IoT devices using Low-Powered Wireless Personal Area Networks (LPWPANs) protocol to defend against various network attacks. \cite{zeitz2017IOTDI} extended the work in \cite{sherburne2014ACISRC} by presenting a design 
based on address rotation to obscure the communications among IoT devices. However, \cite{sherburne2014ACISRC, zeitz2017IOTDI} do not have any experimental validation of the design.  \cite{mahmood2016WFIOT} developed an MTD security framework based on context-aware code partitioning and code diversification for IoT devices to obfuscate the attackers.

\cite{zeitz2017IOTDI} and \cite{zeitz2018WCL} 
developed micro MTD IPv6 as a solution to provide privacy and defense services to resource constrained devices. This work limited the time available to an attacker performing reconnaissance attacks and provided power consumption analysis to prove the efficacy of their micro MTD mechanism.  \cite{Kouachi2018WINCOM} proposed an MTD technique to provide anonymization of packet flow for IoT devices. They looked into privacy issues caused by tracking and communication-flow identification in current low power solutions.  They proposed micro One Time Address that  changes the structure of IPv4 packets and verifies a single address being used only for transmission of one packet.  However, the change of IP header requires re-configuring all the routers. \cite{Nizzi2019IoT} set their goals to provide a lightweight solution to change the addresses of IoT devices. They proposed an MTD solution called HMAC (AShA) performing address shuffling network-wide. A multicast message is sent and the proposed mechanism allows for the devices to recompute their addresses.  \cite{Kahla2018TrustCom} proposed a self-configuring fog architecture to provide security and trust.  The proposed MTD technique moves applications inside a fog network by using live migrations; however, the overhead of live migrations was not considered in their work.

\cite{Wang2019IoT} proposed a game theoretic zero-determinant approach for MTD in IoT to minimize extra operations required by Markov gaming while dominating the game based on a Zero-Determinant (ZD) strategy. \cite{Almohaimeed2019LISAT} proposed a model that prevents attackers from discovering device addresses in IoT networks. Their MTD technique was designed to preserve privacy while transmitting data via a dedicated MTD channel. \cite{Lin2019ICC} proposed security function virtualization MTD to protect SDN enabled smart grid from resource exhaustion attacks.  The proposed MTD migrates virtual security functions upon changes in traffic states.  To minimize the total migration time, they formulated a migration problem as an optimization problem under the scenarios with different network constraints.  \cite{Hamada2018IEMCON} proposed an honeypot-like MTD management framework (called HIoT) to secure an IoT network by deceiving attackers. HIoT uses cell phones around an IoT device to dynamically project these cell phones as fake gateways, real gateways, and sensors.  They also created real and fake sub-nets to carry real and fake data. Similarly, \cite{Vuppala2019GIoTS} proposed an MTD mechanism against side channel attacks. They presented a method to calculate the interval required for encryption re-keying after collecting a minimum number of trace leakages so as to reduce computation overhead. 

Defensive deception techniques provide proactive defense services by adding an extra layer of defense on top of traditional security solutions (e.g., Intrusion Detection System, or IDS, firewalls, or endpoint anti-virus software)~\cite{Miyazaki2014ICNC}.  \cite{La2016IoTJ} introduced a game theoretic method to model the interaction between an attacker who can deceive a defender with suspicious or seemingly normal traffic and a defender in honeypot-enabled IoT networks.  \cite{Anirudh2017ICCCSP} used honeypots for online servers to mitigate Distributed Denial of Service (DDoS) attacks launched from IoT devices.  \cite{Dowling2017ISSC} created a ZigBee honeypot to capture attacks and used it to identify the DDoS attacks and bot malware.  However, none of the works cited above ~\cite{Anirudh2017ICCCSP, Dowling2017ISSC, La2016IoTJ, Miyazaki2014ICNC} analyzed the impact of deception techniques on system-level security. Also, none of the works cited above considered the tradeoff between defense cost vs. system-level security for an IoT system which allows distributed decoy deployment to achieve adequate coverage and provide cost-effective defense service~\cite{DeceptionVendors}.  \cite{cho2018JDMS} investigated an integrated defense system to identify what components of each defense mechanism can provide the best solution for `defense in breadth' considering both enhanced security and defense cost. However, their work is based on model-based analysis without empirical verification. 

All the works cited above focused on either cyberdeception or MTD. There is no current work on developing an integrated defense system equipped with both MTD and defensive deception techniques. Relative to the works cited above, we propose an integrated proactive defense based on cyberdeception and MTD techniques as intrusion preventive mechanisms that can effectively and efficiently mitigate the adverse effect of attackers before the attackers penetrate a target IoT system.

\subsection{Security Models and Metrics}
Graphical security models, including attack graphs (AGs)~\cite{Sheyner2002SP} and attack trees (ATs)~\cite{Saini2008JCSC}, have been widely employed for security analysis in various types of networks.
An attack graph (AG) shows all possible sequences of the attacker actions that eventually reach the target. As the network size increases, the size of an AG can grow exponentially, thus limiting its applicability.  An attack tree (AT) is a tree with nodes representing the attacks and the root representing the goal of attacks. It systematically presents potential attacks in the network.  However, AT is also not scalable with the growth of network size.  

In order to address the scalablity issue, a two-layer Hierarchical Attack Representation Model (HARM) was introduced in ~\cite{Hong2015TDSC} by combining various graphical security models onto different layers.  In a two-layer HARM, the upper layer captures the network reachability information and the lower layer represents the vulnerability information of each node in the network. The layers of the HARM can be constructed independently of each other.  This decreases the computational complexity of calculating and evaluating the HARM compared with that of the existing single-layered graphical security models.
\cite{Ge2017JNCA, Hong2015TDSC} investigated the effectiveness of defense mechanisms based on HARM. In \cite{Ge2017JNCA}, a framework was developed to automate security analysis of an IoT system by which HARM is used to assess the effectiveness of both device-level and network-level defense mechanisms based on various performance metrics such as attack cost and attack impact.  In \cite{Hong2015TDSC}, MTD techniques were evaluated in a virtualized system based on HARM using a risk metric. However, three different MTD techniques, including shuffling, diversity and redundancy, were separately evaluated without considering an integrated defense system. Relative to the works cited above, we also leverage HARM as our graphical security model since it scales with large IoT systems. However unlike the cited works above, we develop a HARM model specifically for security analysis of our proposed integrated defense system using both cyberdeception and MTD techniques.

In the literature, a risk-based 
security model has also been used to assess the effectiveness of defense mechanisms~\cite{Abie2012BAN, Rullo2017ICDCS, Savola2012BAN}. \cite{Abie2012BAN} proposed a risk-based security framework for IoT environments in the eHealth domain to measure expected risk and/or potential benefits by taking a game theoretic approach and context-aware techniques. \cite{Savola2012BAN} proposed an adaptive security management scheme considering security metrics (e.g., metrics representing authentication effectiveness, authorization metrics) to deal with the challenges in eHealth IoT environments. However, 
only high-level ideas about the metrics were described without taking into account key characteristics of IoT environments that would require lightweight solutions. \cite{Rullo2017ICDCS} proposed a method to come up with the optimal security resource allocation plan for an IoT network consisting of mobile nodes using a risk metric estimated by reflecting an economic perspective. However, only device-level evaluations were considered without showing system-level evaluations.
Relative to works cited above,
we do not adopt a risk-based security model for assessing defense mechanisms. Rather, we develop a scalable lightweight HARM model to evaluate the deployment of an integrated defense mechanism for an IoT environment by meeting both system security and performance requirements.

\subsection{SDN Technology for IoT}

Software defined networking (SDN) is a promising technology to flexibly manage complex networks. In the SDN-based architecture, the control logic is decoupled from the switches and routers and implemented in a logically centralized controller; the controller communicates with the data forwarding devices via the southbound application programming interface (API) and provides the programmability of network applications using the northbound API. OpenFlow (OF) is the most widely used southbound API which provides the specifications for the implementation of OF switches (including the OF ports, tables, channels, and protocols)~\cite{openflow2012ONF}.
Some SDN solutions are applied to IoT networks for data flow control among IoT devices~\cite{de2015LAT}, data exchange reduction in wireless sensor networks~\cite{Galluccio2015INFOCOM}, wireless access networks~\cite{Lei2014VITAE}, mobile networks~\cite{Bernardos2014WirelessComm}, smart urban sensing~\cite{Liu2015CommMag}, and topology reconfiguration decision making in wireless sensor networks~\cite{ge2018FGCS}. 
Unlike the above cited works, our work considers a general IoT network with the support of SDN functionality for network topology shuffling where an IoT network consists of both decoy nodes and real nodes.

\section{System Model} \label{sec:system-model}
In this section, we discuss our system model, including (1) the network model in an IoT environment with the support of SDN technology; (2) the attack model describing the attacker's capabilities and attack goals considered in this work; and (3) the defense model addressing defense mechanisms deployed in the given network.

\subsection{Network Model}  \label{sec:network-model}
In this work, we consider an IoT network (e.g., a smart hospital) which consists of servers and IoT nodes. IoT nodes collect data and periodically deliver them to servers via single or multiple hops for further processing. IoT nodes of different functionalities and servers are placed in different Virtual Local Area Networks (VLANs) in the given network. We assume SDN technology~\cite{de2015LAT, Galluccio2015INFOCOM, gartner2003byzantine, Lei2014VITAE} is applied to the IoT network in order to effectively and efficiently manage and control nodes. We consider one SDN controller to be deployed in a remote server. The SDN controller communicates with SDN switches and manages flows between IoT nodes and servers which are connected to switches. Users from the Internet can request services from the servers and will not interact with IoT nodes directly. We will further detail the network scenario in our case study in Section~\ref{ssec:setup}. 

\subsection{Node Model} \label{sec:node-model}
We characterize a node's attributes by four aspects: (1) whether a node is compromised or not (i.e., $n_i.c = 1$ for compromised; $n_i.c = 0$ otherwise); (2) whether a node is a real node or a decoy (i.e., $n_i.d = 1$ for a decoy; $n_i.d = 0$ for a real node); (3) whether a node is a critical node with confidential information that should not be leaked out to unauthorized entities (i.e., $n_i.r = 1$ for a critical node; $n_i.r = 0$ otherwise); and (4) a list of vulnerabilities that a node is vulnerable to (i.e., $n_i.v = \{v_1, ..., v_m\}$ where $m$ is the total number of vulnerabilities). Hence, node $i$'s attributes are represented by:
\begin{equation} \label{eq:node-attributes}
A_{n_i}=[n_i.c,n_i.d,n_i.r,n_i.v].
\end{equation}
    
\subsection{Attack Model}\label{sec:attack-model}
In this work, we consider the following attacks that may lead to breaching system security goals:

\begin{itemize}
\item {\em Reconnaissance attacks}: Outside attackers are able to perform scanning attacks to identify vulnerable targets (e.g., a server) and then break into a system (or a network). The success of this attack demonstrates the successful identification and compromise of vulnerable targets by the outside attacker and leads to the loss of system integrity. This is related to triggering the system failure based on the security failure condition 1 ({\tt SFC1}) in Section \ref{sec:security-fail-conditions}.
\item {\em Data exfiltration attacks}: Inside, legitimate attackers are able to use credentials (e.g., login credentials or a legitimate key to access resources) obtained from a compromised node to leak confidential information to unauthorized, outside entities. The success of this attack results in the leakage of confidential information to unauthorized parties and leads to the loss of confidentiality. This is related to triggering the system failure based on the security failure condition 2 ({\tt SFC2}) in Section \ref{sec:security-fail-conditions}.
\end{itemize}

We make the following {\bf assumptions on attack behaviors and goals} to characterize attackers:
\begin{itemize}
\item An attacker is assumed to have limited knowledge on whether a given node is decoy (i.e., a fake node mimicking a real node) or not. The attacker's capability to detect the deception depends on the knowledge gap between the attacker and the real system state (i.e., how effectively the deployed decoy system mimics the real system in a sophisticated manner). We characterize the level of an attacker's intelligence in detecting a decoy node by the degree (or probability) at which the attacker interacts with the decoy node, as described in Section \ref{sec:defense-model}.
\item An attacker's behavior is monitored after interacting with a decoy. If the attacker realizes the existence of a decoy, it terminates interactions with the decoy immediately and attempts to find a new target to break into the system.
\item An attacker's ultimate goal is to compromise servers to leak confidential information to unauthorized entities outside the IoT network.
\item An attacker is capable of identifying and compromising unpatched exploitable vulnerabilities or unknown vulnerabilities in a given IoT network.
\item An attacker is highly unlikely to compromise servers directly as each server is assumed to have strong protection mechanisms. Therefore, the attacker can exploit vulnerable IoT nodes as entry points, move laterally within the network after the exploitation, and eventually compromise servers by identifying and exploiting unpatched or unknown vulnerabilities.
\item The SDN controller is assumed to be well-protected where communications between the SDN controller and SDN switches are secure~\cite{gartner2003byzantine}.
\end{itemize}

\subsection{Defense Model}\label{sec:defense-model}
We assume traditional defense mechanisms are in place in the IoT network, including a network-based IDS, firewalls, and anti-virus software on servers. The IDS is capable of monitoring the whole IoT network and creates alerts on detected intrusions for incident responses. This work focuses on two types of intrusion prevention mechanisms, namely, cyberdeception and MTD, to divert attackers from real IoT nodes and dynamically change the attack surface to increase attack complexity.

\subsubsection{Decoy System as Defensive Deception}
A defender (i.e., system) can defensively deceive attackers with the purpose of luring them into a decoy system and interacting with them to capture and analyze malicious behaviors and reveal intentions/strategies. The decoy system is deployed independently from the real system. Accordingly, we assume that normal, legitimate users are not aware of the existence of the decoy system while the defender will receive alerts caused by the malicious intrusions if an attacker breaks into the decoy system. We consider two types of decoys utilized throughout an IoT network in this work:
\begin{enumerate}
\item {\bf Emulation-based decoys}: This type of decoys allows defenders to create a variety of fake assets and to provide a large-scale coverage across the network. 
\item {\bf Full OS-based decoys}: This type of decoys enables the replication of actual operating system and software running on production devices to increase the engagement possibility of the attacker. 
\end{enumerate}
Both emulation-based and full OS-based decoys can be autonomously created to fit within the environment without changing the existing infrastructure. To increase overall chances of exploiting decoys by attackers, a combination of diverse forms of decoys with various interactive capabilities can be created to resemble legitimate nodes. 
There exists an intelligence center performing the following tasks: (1) create, deploy, and update a distributed decoy system; (2) provide automated attack analysis, vulnerability assessment, and forensic reporting; and (3) integrate the decoy system with other prevention systems (e.g., security incident and event management platform, firewalls) to block attackers. The module for the decoy node deployment can be implemented and placed in a remote server.

We create a design parameter, $P_d$, indicating the probability that an attacker interacts with a decoy node. To be specific, we consider $P_d^{em}$ as the probability that an attacker interacts with an emulation-based decoy and $P_d^{os}$ as the probability that an attacker interacts with a full OS-based decoy ($P_d^{em} \leq P_d^{os}$ as full-OS-based decoys are considered as having more sophisticated services with more cost).

\subsubsection{Network Topology Shuffling-based MTD}
We consider Network Topology Shuffling-based MTD (NTS-MTD) to change the topology of a given IoT network. NTS-MTD is to be triggered following the concept of event-based MTD in that the network topology changes upon the occurrence of an event. We assume that the SDN controller can control and change flows among nodes in an SDN-based IoT system. We combine cyberdeception and NTS-MTD by means of network topology shuffling to change the attack surface of the IoT network populated with both real and decoy nodes. The details of the proposed decoy system and the event-based NTS-MTD will be described in Section~\ref{sec:proactive-defense}.

\subsection{Security Failure Conditions}\label{sec:security-fail-conditions}
A system fails when either of following two conditions is satisfied:
\begin{itemize}
\item {\bf Security Failure Condition 1 ({\tt SFC1})}: This system failure is closely related to the attacker's successful reconnaissance attacks and accordingly their successful compromise of system components. We define this system failure based on the concept of Byzantine Failure~\cite{gartner2003byzantine}. That is, when more than one third of legitimate nodes are compromised, the system fails due to the loss of system integrity.
\item {\bf Security Failure Condition 2 ({\tt SFC2})}: This system failure occurs when confidential information is leaked out to unauthorized entities by inside attackers (or compromised nodes), which perform data exfiltration attacks. Th system fails due to the loss of data confidentiality.
\end{itemize}

\section{Proposed Proactive Defense Mechanisms}\label{sec:proactive-defense}
In this section, we describe our proposed NTS-MTD technique in three main aspects: (1) when to perform network topology shuffling with decoy nodes; (2) how to perform topology network shuffling with decoy nodes; and (3) graphical security model for security analysis.

\subsection{When to Perform Network Topology Shuffling with Decoy Nodes}\label{ssec:net-topology}
In this section, we describe the initial deployment of decoy nodes in an IoT network and when to perform network topology shuffling with decoy nodes.

\subsubsection{Deployment of Decoy Nodes}
Both server and IoT nodes are deployed in the IoT network. As the network is divided into different virtual local area networks (VLANs), we place IoT decoy nodes into each VLAN based on the deployment of real nodes in the corresponding VLAN. At least one decoy server needs to be deployed to interact with the attacker and reveal the attacker's intent. Note that we can deploy more decoys if the VLAN has a large number of real nodes with different types. When adding decoy nodes, we connect real IoT nodes with decoy nodes to lure attackers into the decoy system. The SDN controller controls flows from real IoT nodes to decoy nodes or from decoy nodes to decoy nodes, and also from real IoT nodes to real IoT nodes. There will be no flows from decoy nodes to real nodes as decoy nodes are used to divert attackers from the real system; once the attacker is lured into the decoy system, it will be diverted to other decoys within the decoy system and the behavior will be monitored; if the attacker detects a decoy node, it will terminate the interaction with the decoy node and look for a new target to break in. Directional flows between real and decoy nodes may reveal some information to attackers in the long term. In this work, We consider changing flows from real nodes to both real and decoy nodes to increase the complexity of connection changes.

Updated flows (either addition or removal) may affect normal flows from IoT nodes to servers for service delivery. In practice, IoT nodes will consume more energy to deliver more flows and may delay the time to send normal packets toward the server. We use packet delivery ratio as a metric for measuring service availability, as discussed in Section~\ref{ssec:metrics}.

Initially we create decoy nodes with added connections to some randomly chosen real nodes based on the deployment of real nodes in each VLAN. The randomly generated network topology will be used as the initial topology and then fed into the shuffling optimization algorithm to identify an optimal network topology.

\subsubsection{When to Shuffle Network Topology}\label{ssec:adaptive-shuffling}
We can use a fixed time interval to execute NTS-MTD. Apparently the fixed time interval is the most important parameter of this strategy because if the interval is too short, the defense cost will be high although the system lifetime may be prolonged because frequent topology shuffling can mislead the attacker to decoy paths and nodes and thus keep real nodes from the attacker. On the other hand if the fixed time interval is too long, it will adversely shorten the system lifetime because of infrequent network shuffling. We call this strategy the fixed time interval strategy or just ``fixed'' for short. A variation of this ``fixed'' strategy is to have the time interval follow a distribution with the mean being the same as the fixed time interval used by the fixed strategy. This will add some stochastic nature to the time interval which is treated as a random variable. We will call this strategy ``random'' for short.

Alternatively, we can execute NTS-MTD when a condition is detected true, for example, based on the system security vulnerability level detected by the system. We will call this strategy ``adapative'' for short. To be specific, the system security vulnerability level at time $t$, denoted by $SSV(t)$, is measured by two dimensions: (1) how many legitimate, inside nodes are compromised until time $t$, which is associated with {\tt SFC1}; and (2) how many neighboring nodes of a critical node (i.e., $n_i.r = 1$) within $k$ hops from the critical node $i$ are compromised until time $t$, which is related to {\tt SFC2}. Of course we do not know which node is actually compromised unless the IDS has detected it. However, given the list of vulnerabilities that a node is vulnerable to, as discussed in Section~\ref{sec:node-model} and the compromise rate for each vulnerability which is documented in several sources such as \cite{nist2005NVD} we can estimate the probability that a node is compromised at time $t$. Note that when the system meets either {\tt SFC1} or {\tt SFC2}, the system fails, leading to $SSV(t) = 1$. Otherwise, $SSV(t)$ is computed by:
\begin{equation} \label{eq:ssv}
SSV(t) = w_1 \frac{CN(t)}{N} + w_2 \frac{CN_{ck} (t)}{N_{ck} (t)}
\end{equation}
Here $w_1$ and $w_2$ are weights to consider {\tt SFC1} and {\tt SFC2}, respectively, where $w_1 + w_2 = 1$. $N$ is the total number of real nodes which is known at deployment time and $CN(t)$ is the number of compromised, real nodes at time $t$ which may be estimated from the compromise rate of each vulnerability that a node is vulnerable to. (See more about this in Section~\ref{ssec:setup}.) $N_{ck}(t)$ is the total number of real nodes within $k$ hops from given critical nodes at time $t$ which may be obtained from the topology shuffled at time $t$ while $CN_{ck}(t)$ is the total number of compromised, real nodes within $k$ hops from critical nodes which again can be estimated from the compromise rate of each vulnerability that a node is vulnerable to. Since there may be multiple critical nodes which have confidential information that should not be leaked to outside unauthorized parties, we estimate $CN_{ck}(t)$ by:
\begin{equation} \label{eq:eqn2}
CN_{ck}(t)=\sum_{i\in L_{k}(t)} n_i.c(t)
\end{equation}
where $L_k(t)$ is the number of real nodes that belong to neighbors of any critical nodes within $k$ hops from them at time $t$ and $n_i.c(t)$ refers to whether node $i$ is compromised $(n_i.c(t)=1)$ or not $(n_i.c(t)=0)$ at time $t$. The cardinality of $L_k(t)$ (i.e., $|L_k(t)|$) yields $N_{ck}(t)$. Note that as the network topology keeps changing due to the execution of NTS-MTD, both $N_{ck} (t)$ and $CN_{ck} (t)$ are functions of time to reflect their dynamic changes. If $L_k(t)$ includes any critical nodes being compromised, the system meets {\tt SFC2} and fails. That is, $SSV(t)=1$ and no further detection of system security level is needed. 

Lastly we can have a hybrid strategy that will degenerate to the adaptive strategy when the triggering time as determined by the adaptive strategy is smaller than the fixed time interval used by the fixed strategy and will degenerate to the fixed strategy otherwise.

The four ``when-to-shuffle'' strategies will be more formally defined and labeled later in Section \ref{ssec:comparing-schemes} when we perform evaluation.


\subsection{How to Shuffle Network Topology with Decoy Nodes}\label{ssec:net-shuffling}
We develop three strategies to address how to perform network shuffling when it is time to execute NTS-MTD. The basic idea of our design is to maximize the chance of the attacker exploiting decoy targets, thus effectively deterring or preventing its security attacks to real nodes. In order to reach a target node, an attacker could exploit a node as an entry point and use it as the stepping stone to compromise other nodes and further compromise the target. It may be able to find multiple attack paths via one or multiple entry points. An attack path describes a sequence of nodes that an attacker could compromise to reach the target node. We consider a set of attack paths $AP$ for an attacker to reach all targets from all possible entry points. Each attack path $ap$ is a sequence of nodes along the path. We use $AP_r$ to represent a set of attack paths with real nodes as targets and $AP_d$ to denote a set of attack paths with decoy nodes as targets. $AP_r$ only contains real nodes while $AP_d$ contains both real and decoy nodes. To be specific, if an attacker finds a real node as the entry point and compromises other real nodes until reaching a real target node, this is counted as an attack path in $AP_r$; however, it could be diverted to a decoy node. Once the attacker is lured into the decoy system, it will be diverted to other decoy nodes within the decoy system. If the attacker reaches a decoy target node, this is counted as an attack path in $AP_d$; however, if the attacker figures out the decoy node and terminates its interaction, it is not counted as an attack path because the attacker does not reach the decoy target node. Besides, decoy nodes could be updated or cleared once it is detected compromised by the intelligence center in which case the attacker will not recognize the same decoy node during subsequent attacks.

To maximize the chance of the attacker being misled to decoy targets, we develop the following two ``how-to-shuffle'' strategies: 
\begin{itemize}
\item \textbf{GA-based optimization:} We design three metrics to be optimized in the algorithm: (1) The number of attack paths toward the decoy targets ($N_{DT}^{AP}$); (2) Mean Time To Security Failure (MTTSF); and (3) Defense cost ($C_D$). Computations of these metrics are described in Section~\ref{ssec:metrics}.

\item \textbf{Decoy path-based optimization:} Due to the high computational complexity of GA, we design a simple heuristic algorithm to provide a close-to optimal solution in topology shuffling. The algorithm takes a path-based optimization approach in two ways: (i) Shuffle edges (connections) from real IoT nodes to decoy nodes to randomize decoy connections; and (ii) Shuffle edges (connections) among real IoT nodes to maximize the number of attack paths toward decoy targets.
\end{itemize}

The third strategy is a baseline strategy that generates a network topology based on a connection probability of a real/decoy node being connected to another decoy node. We call this strategy ``random'' meaning that the connection probability is a random variable in the range of [0, 1] which determines if a connection from a real/decoy node to another decoy node should be created in the resulting topology.

The three ``how-to-shuffle'' strategies will be more formally defined and labeled later in Section \ref{ssec:comparing-schemes} when we perform evaluation.

\subsection{Graphical Security Model for Security Analysis of NTS-MTD}\label{ssec:graph-security-model}

We develop a graphical security model based on HARM to assess the security of an IoT network. 

\begin{figure*}[htbp]
\vspace{-5mm}
    \centering
    \includegraphics[width=0.9\textwidth]{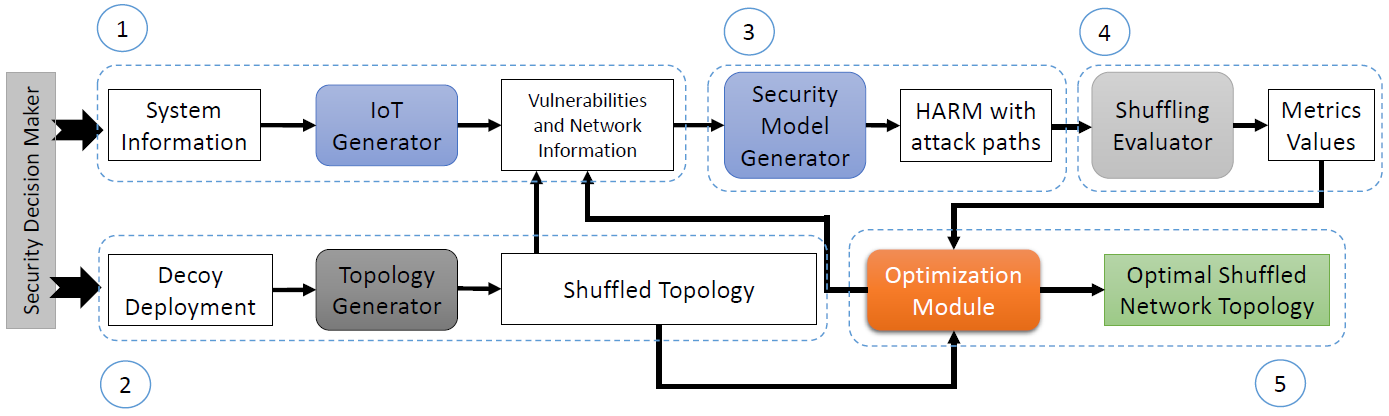}
\caption{Workflow for the security analysis.}
\label{fig:workflow-analysis}
\end{figure*} 
Fig.~\ref{fig:workflow-analysis} describes the workflow of our security analysis in five phases: network generation, topology generation, security model generation, shuffling mechanism evaluation, and shuffling optimization. 

\begin{enumerate}
   \item {\bf Phase 1}: The security decision maker provides the {\tt IoT Generator} with the system information (i.e., an initial network topology and node vulnerability) to construct an IoT network.
   \item {\bf Phase 2}: Given the network and initial deployment of decoys, the {\tt Topology Generator} randomly generates a set of different topologies for GA-based shuffling and one topology for decoy path-based shuffling (i.e., add connections from real nodes to decoys/real nodes). 
   \item {\bf Phase 3}: The {\tt Security Model Generator} takes the shuffled network as input and automatically generates a HARM model that captures all possible attack paths. We use a three-layer HARM
   as our graphical security model, with the upper layer capturing the subnet reachability information, the middle layer representing the node connectivity information (i.e., nodes connected in the topological structure), and the lower layer denoting the vulnerability information of each node. 
   \item {\bf Phase 4}: The {\tt Shuffling Evaluator} takes the HARM model as input along with evaluation metrics and computes results which are then fed into the {\tt Optimization Module}. 
   \item {\bf Phase 5}: For GA-based shuffling, based on the initial set of shuffled topologies and associated evaluation results, the {\tt Optimization Module} applies the multi-objective GA to compute the optimal topology for the IoT network.
   For decoy path-based shuffling, the {\tt Optimization Module} takes the randomly shuffled topology from the {\tt Topology Generator} and runs the heuristic algorithm to compute the close-to optimal topology.
\end{enumerate}

\section{Numerical Results \& Analysis}\label{sec:results}
In this section, we first describe the simulation setup, performance metrics used for performance analysis, parameter table, implementation detail, and data collection process. Then we conduct a comparative performance analysis of 12 schemes of when and how to execute our proposed NTS-MTD technique.

\begin{figure}
    \centering
    \includegraphics[width=0.4\textwidth]{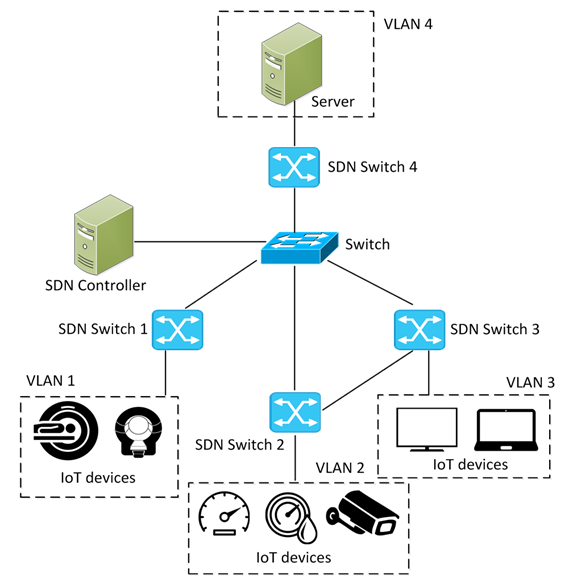}
\caption{A software-defined IoT network.}
\label{fig:iot-example}
\end{figure}

\subsection{Simulation Setup}\label{ssec:setup}
We use an IoT network shown in Fig.~\ref{fig:iot-example} in our simulation and assume SDN is deployed to support connection changes. We consider a smart hospital scenario in the IoT context. Specifically, the network consists of four VLANs. There are two Internet of Medical Things (i.e., MRI and CT Scan) in VLAN1 (e.g., medical examination rooms), a smart thermostat, a smart meter, and a smart camera in VLAN2 (e.g., medical care units), a smart TV and a laptop in VLAN3 (e.g., staff office) and a server located in VLAN4 (e.g., server room). At the initial deployment, VLAN4 is connected with other three VLANs as IoT devices need to deliver information to the server for further processing. VLAN2 is also connected to VLAN3 for applications running on the laptop to control smart sensors as well as receive videos from the smart camera.

\begin{table*}[h]
\renewcommand{\arraystretch}{1.2}
\caption{Real node and vulnerability information.}
    \centering
    \begin{tabular}{|P{3cm}|P{3cm}|P{3cm}|P{3cm}|}
    \hline
    {Real Node} & VLAN  & CVE ID & {Compromise Rate}\\
    \hline
    MRI   & VLAN1 & CVE-2018-8308 & 0.006 \\
    \hline
    CT Scan & VLAN1 & CVE-2018-8308 & 0.006 \\
    \hline
    Smart Thermostat & VLAN2 & CVE-2018-11315 & 0.006 \\
    \hline
    Smart Meter & VLAN2 & CVE-2017-9944 & 0.042 \\
    \hline
    Smart Camera & VLAN2 & CVE-2018-10660 & 0.042 \\
    \hline
    Smart TV & VLAN3 & CVE-2018-4094 & 0.012 \\
    \hline
    Laptop & VLAN3 & CVE-2018-8345 & 0.004 \\
    \hline
    Server & VLAN4 & CVE-2018-8273 & 0.006 \\
    \hline
    \end{tabular}
    \label{tab:node_vulnerability}
\end{table*}

\begin{table*}[th!]
\caption{\sc Decoy Node and Vulnerability Information.}
\renewcommand{\arraystretch}{1.2}
\label{tab:Decoy_vulnr}
    \centering
\begin{tabular}{|P{3cm}|P{3cm}|P{3cm}|P{3cm}|}
        \hline
    {Decoy Node} & VLAN & CVE ID & {Compromise Rate} \\
    \hline
    \multirow{2}{*}{CT Scan} & \multirow{2}{*}{VLAN1} & CVE-2018-8308 & 0.006 \\
    \cline{3-4}          &       & CVE-2018-8136 & 0.012 \\
    \hline
    \multirow{3}{*}{Smart Camera} & \multirow{3}{*}{VLAN2} & CVE-2018-6294 & 0.042 \\
    \cline{3-4}          &       & CVE-2018-6295 & 0.042 \\
    \cline{3-4}          &       & CVE-2018-6297 & 0.042 \\
    \hline
    \multirow{2}{*}{Smart TV} & \multirow{2}{*}{VLAN3} & CVE-2018-4094 & 0.012 \\
    \cline{3-4}          &       & CVE-2018-4095 & 0.012 \\
    \hline
    \multirow{3}{*}{Server} & \multirow{3}{*}{VLAN4} & CVE-2016-1930 & 0.042 \\
    \cline{3-4}          &       & CVE-2016-1935 & 0.012 \\
    \cline{3-4}          &       & CVE-2016-1962 & 0.042 \\
    \hline
    \end{tabular}
\end{table*}

We collect software vulnerabilities from Common Vulnerabilities and Exposures (CVE)/National Vulnerability Database (NVD) \cite{nist2005NVD}. We assume each real node has one vulnerability that could be exploited by the attacker to gain a root privilege. More vulnerabilities could be chosen for nodes in the future work. This research work focuses on proposing and evaluating the integrated proactive defense mechanism, rather than demonstrating capabilities of the graphical security model to analyze the security posture of the IoT network with multiple vulnerabilities. The vulnerability information of real nodes (i.e., CVE ID) is presented in Table~\ref{tab:node_vulnerability}. We also assume the compromise rate of each vulnerability. The compromise rate represents the frequency that an attacker could successfully exploit the vulnerability to gain root privilege per time unit (i.e., hour). We estimate the value according to the base score from the Common Vulnerability Scoring System (CVSS). Specifically, we estimate the compromise rate as once per day (i.e., 0.042) if the base score is 10.0, twice per week (i.e., 0.012) if the base score of is around 8.0, once per week (i.e., 0.006) if the score is around 7.0, and once per 10 days (i.e., 0.004) if the score is around 5.0. This value will be used to calculate Mean Time to Compromise (MTTC) and Mean Time to Security Failure (MTTSF) by the HARM model. Once a node is compromised it can perform packet dropping or manipulating attacks to affect service availability. In practice, however, a compromised node may not drop or manipulate a packet passing through it, so it won't get caught by the network IDS. In our simulation, we consider a packet drop probability $P_a^d$ and a packet manipulation probability $P_a^m$ by the attacker. 

We put one decoy node in each VLAN in the initial deployment of the decoy system. In order to lure attackers, each decoy is assumed to be configured to have multiple vulnerabilities. An attacker could exploit any vulnerability to gain the root permission of the node. The vulnerability information of decoys is listed in Table~\ref{tab:Decoy_vulnr}. We use emulated decoys for the CT scan, smart camera, smart TV, and full-OS based server.

\subsection{Metrics}\label{ssec:metrics}
We use the following metrics to measure security, performance, and service availability of the proposed proactive defense mechanisms:
\begin{itemize}
\item {\bf Number of attack paths toward decoy targets} ($N_{DN}^{AP}$): This metric indicates the level of deception that diverts an attacker from the real system. $N_{DN}^{AP}$ is calculated by $|AP_d|$ to sum attack paths toward the decoy targets. 
\item {\bf Mean Time To Compromise (MTTC)}: This metric refers to the total amount of time that an attacker takes to compromise a series of nodes within the network until the system reaches a certain security vulnerability level $SSV$. MTTC is estimated by:
\begin{equation} \label{eq:eqn4}
MTTC = \sum_{i \in S} S_i \int_{t=0}^\infty P_i(t)dt
\end{equation}
where $S$ refers to a set of all system states and $S_i$ is 1 when in state $i$ the system does not reach the given $SSV$ level and is 0 otherwise. $P_i(t)$ is the probability of the system being in state $i$ at time $t$.
\item {\bf Mean Time To Security Failure (MTTSF)}: This metric measures the system lifetime indicating how long the system prolongs until the system reaches either {\tt SFC1} or {\tt SFC2} (described in Section \ref{sec:security-fail-conditions}). That is, MTTSF measures the system lifetime without occurring any security failure. MTTSF is measured by:
\begin{equation} \label{eq:eqn3}
MTTSF = \sum_{i \in S} (1-SF_i) \int_{t=0}^\infty  P_i(t)dt
\end{equation}
where $S$ is a set of all system states and $SF_i$ returns 1 when system state $i$ reaches either {\tt SFC1} or {\tt SFC2}; 0 otherwise. $P_i(t)$ indicates the probability of the system being in state $i$ at time $t$.
\item {\bf Defense Cost} ($C_D$): This metric depicts the cost associated with shuffling operations. That is, we count the number of edges shuffled (i.e., from connected to disconnected or from disconnected to connected) by:
\begin{equation} \label{eq:eqn5}
C_D = \int_{t=0}^{MTTSF} C_S(t)
\end{equation}
where $C_S(t)$ refers to the number of shuffled edges at time $t$. Note that the same edge can be shuffled multiple times over time and each shuffling is counted as a separate MTD operation during the system uptime.
\item {\bf Packet Delivery Ratio (PDR)}: This metric measures service availability affected by topology shuffling. Because of topology shuffling, attackers tend to compromise nodes on attack paths. For each attack path in $AP_r$, a compromised node along the path may drop or manipulate packets travelling through it, thereby affecting service availability for service packets passing through the attack path.
If packets are not dropped or manipulated by compromised nodes along the path (because the attacker may not want to get caught by the IDS) or if there is no compromised node along the attack path, then the path will be able to successfully deliver service packets. At each shuffling operation, we count the number of attack paths that can successfully do packet delivery and divide it by the total number of attack paths $|AP_r|$. When the system reaches either {\tt SFC1} or {\tt SFC2}, we calculate the mean PDR over all shuffling operations. 
\end{itemize}

\subsection{Twelve Schemes to Execute NTS-MTD based on When and How Strategies} \label{ssec:comparing-schemes}
We investigate two aspects of NTS-MTD: (i) when-to-shuffle a network topology (in an interval or in an adaptive manner); and (ii) how to select a network topology (by a GA-based optimization, a decoy path-based optimization, or random shuffling). 

Four strategies regarding \textbf{when-to-shuffle a network topology} are:
\begin{itemize}
\item \textbf{Fixed Shuffling (FS)}: This strategy is to execute NTS-MTD in a fixed time interval, $\gamma_1$, to shuffle the network topology.
\item \textbf{Random Shuffling (RS)}: This strategy is to execute NTS-MTD in a random interval based on exponential distribution with mean $\lambda$.
\item \textbf{Adaptive Shuffling (AS)}: This strategy is to execute NTS-MTD in an \textit{adaptive manner} based on $SSV(t)$ with two given thresholds: (1) $\beta$ to check the decrease of the $SSV$ during a checking interval $\Delta$; and (2) $\rho$ to check the current system security vulnerability, $SSV(t)$, as described in Section~\ref{ssec:net-topology}. NTS-MTD is executed when the condition, $(SSV(t)-SSV(t-\Delta)>\beta) \land (SSV(t)>\rho)$, is true. This condition is checked whenever the system detects a compromised real node, thus reflecting the nature of an event-driven adaptive MTD. 
\item \textbf{Hybrid Shuffling (HS)}: This strategy is a mixture of AS and FS. 
Since AS triggers the execution of NTS-MTD until the event condition is detected, it may delay the execution of NTS-MTD unnecessarily especially in the beginning because security vulnerability does not necessarily increase rapidly in the beginning. 
To remedy this, we introduce an upper bound time limit (i.e., the maximum delay) for NTS-MTD execution. Specifically, the time interval to execute NTS-MTD is set to $\mathrm{min}[Int(AS), \gamma_2]$ where $Int(AS)$ returns a time interval when AS is used and $\gamma_2$ is the fixed time interval for the maximum delay when FS is used.
\end{itemize}

Three strategies regarding \textbf{how to select a network topology} are:
\begin{itemize}
\item \textbf{Random Network Topology (RNT)}: This strategy is a baseline strategy that selects a network topology based on a rewiring probability $P_r$ of a node being connected with another node. Here $P_r$ is critical in determining the overall network density in a given network.
\item \textbf{GA-based Network Topology (GANT)}: This strategy selects a network topology that maximizes objective functions used in the GA, as discussed in Section~\ref{ssec:net-topology}.
\item \textbf{Decoy Path-optimized Network Topology (DPNT)}: This strategy selects a network topology that maximizes the number of decoy paths for each real IoT node, as discussed in Section~\ref{ssec:net-topology}. 
\end{itemize}

Since we have four ``when'' strategies and three ``how'' strategies for NTS-MTD execution, there are 12 schemes resulting from the combination of one ``when'' strategy and one ``how'' strategy, viz., FS-RNT (i.e., execution of NTS-MTD based on Fixed Shuffling (FS) and Random Network Topology (RNT)), RS-RNT, AS-RNT, HS-RNT, FS-GANT, RS-GANT, AS-GANT, HS-GANT, FS-DPNT, RS-DPNT, AS-DPNT, and HS-DPNT.

\subsection{Parameter Table, Implementation Detail, and Data Collection Process}
\begin{table*}[ht]
\caption{Design parameters, their meanings and default values.}
    \centering
    \begin{tabular}{|P{1cm}|p{10cm}|P{1cm}|}
\hline
    Param. & \multicolumn{1}{c|}{Meaning} & Value\\
\hline
    $w_1$ & A weight to consider the security vulnerability associated with {\tt SFC1} & 0.5 \\
\hline
    $w_2$ & A weight to consider the security vulnerability associated with {\tt SFC2} & 0.5 \\
\hline
    $P_d^{em}$ & Interaction probability of an attacker with an emulated decoy & 0.9 \\
\hline
    $P_d^{os}$ & Interaction probability of an attacker with a full-OS based decoy & 1.0 \\
\hline
    $P_a^d$ & Probability of a packet to be dropped & 0.5 \\
\hline
    $P_a^m$ & Probability of a packet to be manipulated & 0.5 \\
\hline
    $k$ & Number of hops to determine a node's ego network & 1 \\
\hline
    $N$ & Total number of network topologies with initial decoy deployment and randomly generated connections between real and decoy nodes used in GANT & 100\\
\hline
    $w_N$ & A weight to consider in objective function used in GANT & 1/3 \\
\hline
    $w_M$ & A weight to consider in objective function used in GANT & 1/3 \\
\hline
    $w_C$ & A weight to consider in objective function used in GANT & 1/3 \\
\hline
    $N_g$ & Maximum number of the generation used in GANT & 100 \\
\hline
    $r_c$ & Crossover rate used in GANT & 0.8 \\
\hline
    $r_m$ & Mutation rate used in GANT & 0.2 \\
\hline
    $P_r$ & Probability of an edge being shuffled in RNT (i.e., add/remove an edge) & 0.5 \\
\hline
    $\beta$ & Threshold used to estimate the decrease of the system security vulnerability level during the time used in AS/HS & 0.01 \\
\hline
    $\rho$ & Threshold of tolerating system security vulnerability used in AS/HS & 0.1 \\
\hline
    $\gamma_1$ & Fixed shuffling time interval used in FS (hour) & 24 \\
\hline
    $\gamma_2$ & Fixed shuffling time interval (maximum delay) used in HS (hour) & 120 \\
\hline
    $\lambda$ & Mean value used for exponential distribution in RS (hour) & 24 \\
\hline
    \end{tabular}
    \label{tab:parameters}
\end{table*}

Table~\ref{tab:parameters} summarizes the model parameters, their meanings, and default values used in our simulation runs.

Our proposed NTS-MTD technique is implemented based on the workflow shown in Fig.~\ref{fig:workflow-analysis}. The {\tt Optimization Module} implements the algorithms to execute the three ``how'' strategies, i.e., RNT, GANT, and DPNT, as discussed in Section~\ref{ssec:comparing-schemes}.  

We assume there is an attacker exploiting node vulnerabilities. In each simulation run, the attacker will randomly choose entry points and compromise nodes along the attack paths with behaviors defined in Section \ref{sec:attack-model} until either {\tt SFC1} or {\tt SFC2} (see Section \ref{sec:security-fail-conditions}) is met. We assume that the system will clear decoy nodes once it detects the attacker's interaction with the decoy target. Therefore, the attacker will not recognize the same decoy node in its subsequent action. Decoy nodes are also cleared at each shuffling. By using FS/RS strategies, the network may be shuffled periodically or randomly right at the moment a node is under attack. We assume that the attacker is forced to quit the network due to lost connections and needs to find other ways to break into the network. In the subsequent attack after shuffling, the attacker could continue its previous attack action once it encounters the same real node next time (i.e., MTTC for the real node is accumulated throughout the MTTSF). By using the AS strategy, the network is shuffled due to changes to $SSV$ being detected by the defender. The attacker is also forced to quit the network after each shuffling due to lost connections and needs to find ways to re-enter the network. 
Each newly shuffled network is modeled by a HARM model for calculating potential attack paths. The attacker's intelligence, estimated by $P_d^{em}$ and $P_d^{OS}$ (see Section \ref{sec:defense-model}), is incorporated into the calculation of MTTSF as well as MTTC. We encode each shuffling solution for the whole network as a binary valued vector with 1 representing the existence of an edge between two nodes and 0 representing no edge. We limit potential connections to be edges from real IoT nodes to either decoy nodes or real IoT nodes. Hence, to optimize the defense cost, we aim to maximize $C_T(t)-C_D(t)$ where $C_T(t)$ refers to the total defense cost (i.e., the total number of potential edge changes at time $t$) and $C_D(t)$ is the number of edges changed by executing NTS-MTD at time $t$ (see Section~\ref{ssec:metrics}). 

For GANT, we aim to solve a multi-objective optimization (MOO) problem with three objectives to maximize $N_{DN}^{AP}$ and MTTSF while minimizing $C_D$ (or maximizing $C_T(t)-C_D(t)$). The optimization problem is to compute a set of Pareto optimal solutions (or Pareto frontier)~\cite{cho2017CST}. In order to choose one optimal solution among the Pareto frontier, we first normalize three metrics, denoted by $\widetilde{N_{DN}^{AP}}$, $\widetilde{MTTSF}$ and, $\widetilde{C_D}$, and then assign a weight to each metric based on scalarization-based MOO technique to transform the MOO problem to a single-objective optimization (SOO) problem~\cite{Cho17}. The normalized metric, $\Tilde{X}$, is given by:
\begin{equation} \label{eq:eqn6}
\Tilde{X}=\frac{X}{X_{max}}
\end{equation}
where $X$ is the original metric value and $X_{max}$ is the maximum metric value of the corresponding fitness function in the final population in the GA-based algorithm.

The objective function we aim to maximize is represented by:
\begin{equation} \label{eq:eqn7}
\max \; \;  w_N \widetilde{N_{DN}^{AP}} + w_M \widetilde{MTTSF} + w_C \widetilde{C_D}
\end{equation}
where $w_N$, $w_M$, and $w_C$ are weights to the three metrics with $w_N+w_M+w_C=1$. The optimal solution is the network topology with the maximum objective value. 

In each simulation run, we collect data to calculate the mean time to security failure, MTTSF, 
the number of attack paths toward decoy targets, $N_{DT}^{AP}$, 
the defense cost per time unit, $C_D$, 
and the packet delivery ratio, PDR.
We run the simulation 100 times using random seeds in each simulation. 
After 100 simulation runs, we collect the means of MTTF, $N_{DT}^{AP}$, $C_D$, and PDR
for performance analysis.

\subsection{Comparative Performance Analysis} \label{ssec:performance-analysis}
In this section, we conduct a comparative performance analysis of the 12 schemes discussed in Section~\ref{ssec:comparing-schemes}. We follow the parameter table in Table~\ref{tab:parameters}. We vary the level of attackers' intelligence in detecting decoy nodes (i.e., $P_d^{em}$ and $P_d^{os}$), attack severity (i.e., packet drop probability $P_a^d$ and packet manipulation probability $P_a^m$), the number of decoys in each VLAN, and the number of real IoT nodes to analyze their effects on performance in terms of the mean time to security failure, MTTSF, 
the number of attack paths toward decoy targets, $N_{DT}^{AP}$, 
the defense cost per time unit, $C_D$, 
and the packet delivery ratio, PDR.

\subsubsection{Comparison of Schemes under the Baseline Scenario}\label{sssec:compare-scheme}

We first consider a baseline scenario in which there is only one decoy in each VLAN and the attacker intelligence is low characterized by its high interaction probabilities with decoys, i.e., $P_d^{em}$=0.9 for an emulated decoy and $P_d^{os}$=1.0 for a full-OS based decoy. Recall that a high interaction probability means that the attacker must interact with a decoy node intensively in order to detect it is a decoy.

\begin{figure*}
  \centering
  \subfloat[$N_{DT}^{AP}$ ]{\includegraphics[width=0.5\textwidth]{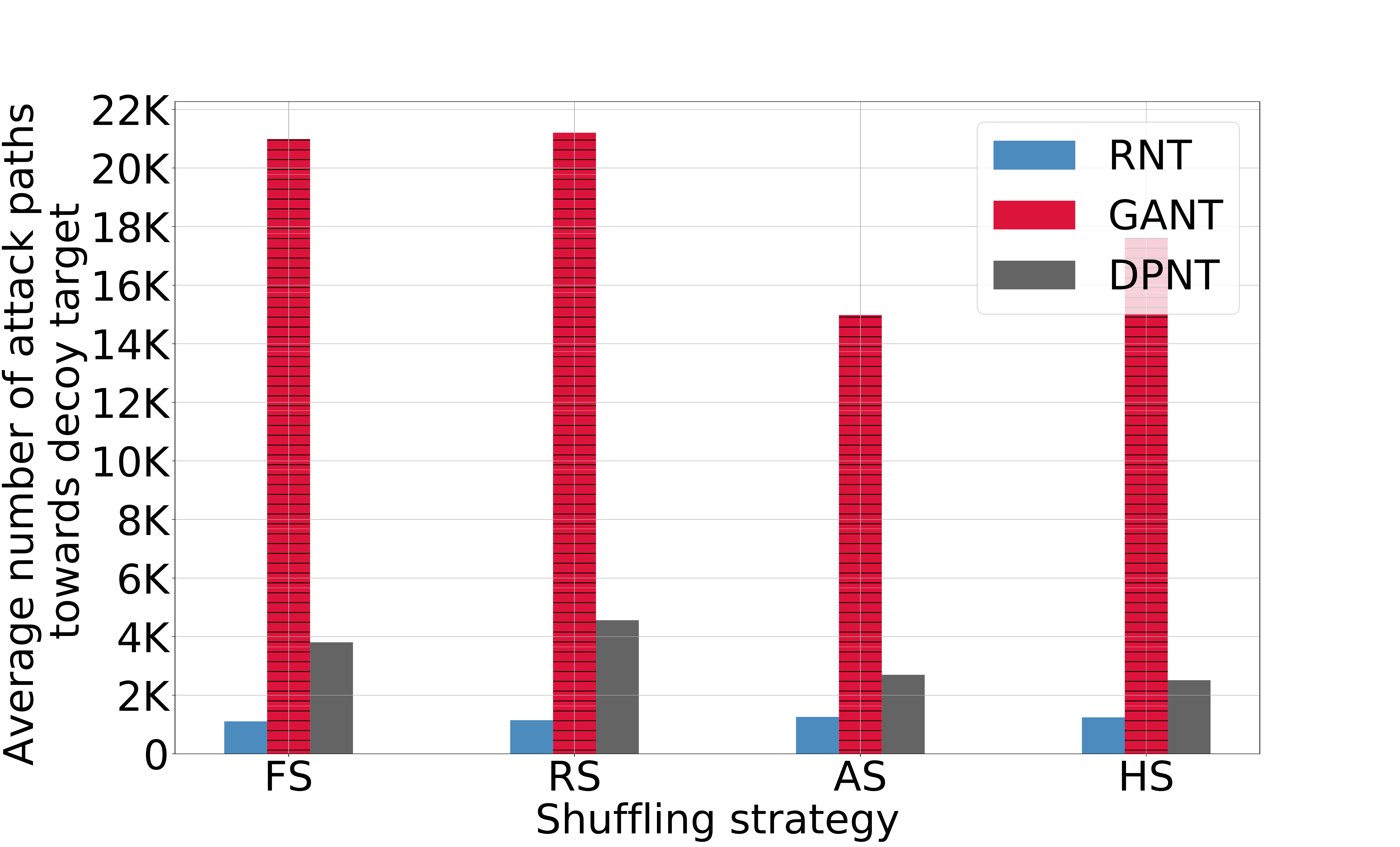}\label{fig:base_ap}}
  \hfill
  \subfloat[MTTSF]{\includegraphics[width=0.5\textwidth]{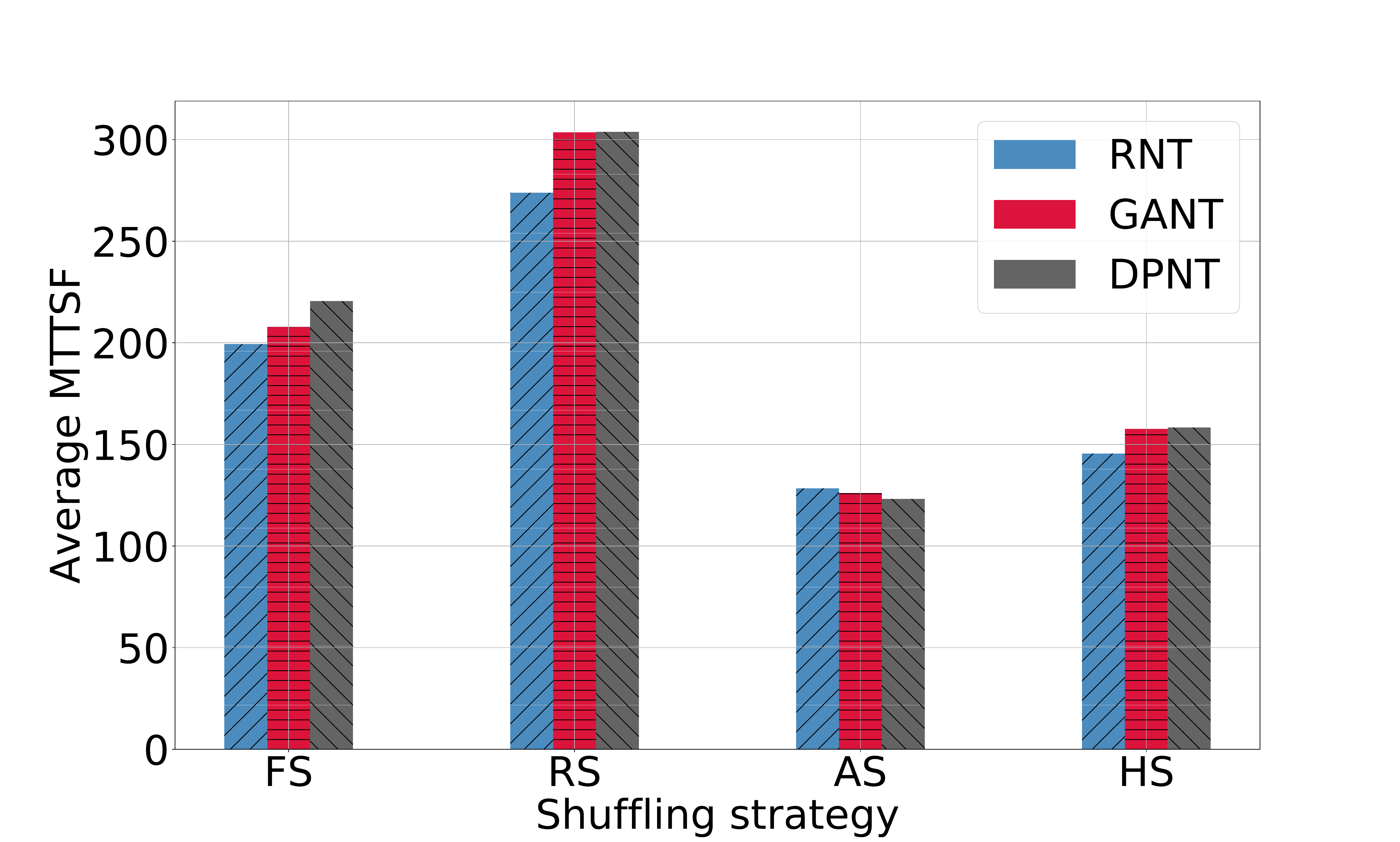}\label{fig:base_mttsf}}
  \hfill
  \subfloat[$C_D$]{\includegraphics[width=0.5\textwidth]{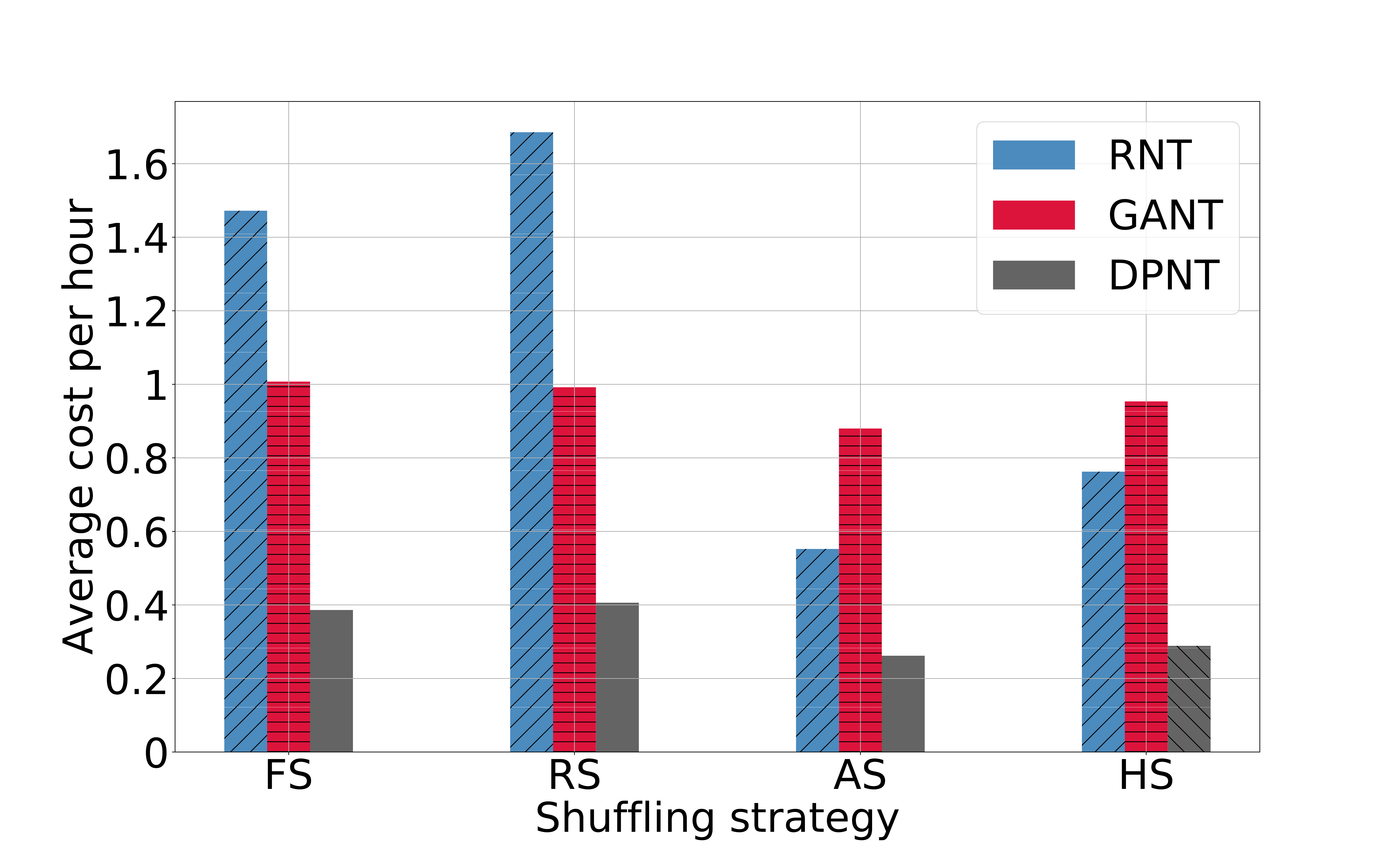}\label{fig:base_cost}}
  \hfill
  \subfloat[PDR]{\includegraphics[width=0.5\textwidth]{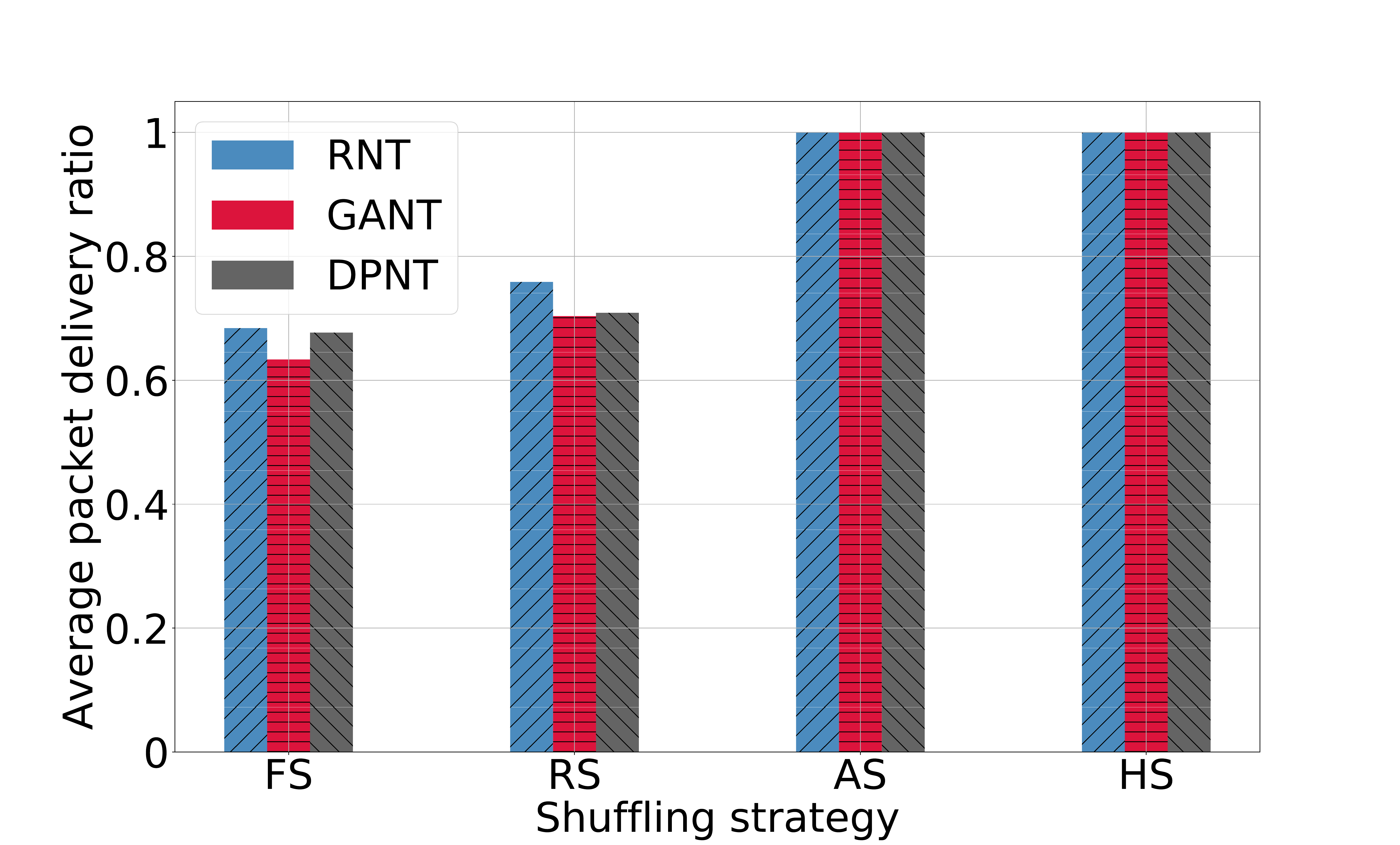}\label{fig:base_ratio}}
  \caption{Comparison of schemes under the baseline scenario.}\label{fig:base-scenario}
\end{figure*}

Fig.~\ref{fig:base-scenario} compares the performance characteristics of the 12 schemes discussed in Section~\ref{ssec:comparing-schemes} for executing our proposed NTS-MTD technique under the default parameters presented in Table~\ref{tab:parameters}. 
\begin{itemize}
\item 
Fig.~\ref{fig:base_ap} compares the number of attack paths toward decoy targets $N_{DT}^{AP}$ (the higher the better) representing deception effectiveness. In the ``when-to-shuffle'' category, fixed/random shuffling (FS/RS) based schemes perform comparably among themselves 
On the other hand, in the ``how-to-shuffle'' category, the genetic algorithm network topology (GANT) scheme performs the best in deception effectiveness, followed by the decoy path-optimized network topology (DPNT) scheme and the random network topology (RNT) scheme. This indicates how-to-shuffle the network has a major impact on deception effectiveness. 
\item
Fig.~\ref{fig:base_mttsf} compares MTTSF (the higher the better) representing the system lifetime before the system experiences a failure.
In the ``when-to-shuffle'' category, fixed/random shuffling (FS/RS) based schemes significantly outperform adaptive/hybrid shuffling (AS/HS) based schemes in MTTSF. One factor is system failure in AS/HS is determined by either {\tt SFC1} or {\tt SFC2} being triggered, or SSV exceeding the threshold. In the current setting, AS/HS uses a low SSV threshold (i.e., 0.1), which indicates a low tolerance on {\tt SFC1} and {\tt SFC2}. This means system status could be considered as failure based on SSV threshold before either {\tt SFC1} or {\tt SFC2} is triggered.
Another factor is that in FS/RS based schemes, a node may be under attacks while the network is shuffled because the fixed/random interval for topology shuffling could be much smaller than the MTTC of the node at which time topology shuffling is triggered by AS/HS. After each shuffling, the attacker is forced to quit the network due to lost connections and needs to re-enter the network by randomly choosing entry points to compromise. After re-entering, the attacker could continue its previous attack once it encounters the same real node next time or launch a new attack for a decoy node as decoys are cleared at each shuffling. This could effectively lead to an increase of MTTSF over time in order to meet either {\tt SFC1} or {\tt SFC2} security failure condition. We see that RS produces the highest MTTSF among all.
In the ``how-to-shuffle'' category, 
GANT and DPNT perform comparably among themselves and both outperform RNT. 
\item
Fig.~\ref{fig:base_cost} compares the defense cost $C_D$ (the lower the better). Since the defense cost is inversely related to the number of attack paths toward decoy targets (i.e., deception effectiveness), we expect the trend for defense cost is just opposite to that in Fig.~\ref{fig:base_ap} for deception effectiveness. This is indeed the case. In the ``when-to-shuffle'' category, adaptive/hybrid shuffling (AS/HS) based schemes perform comparably among themselves and outperform fixed/random shuffling (FS/RS) based schemes, a trend that is opposite to that for deception effectiveness. In the ``how-to-shuffle'' category, DPNT performs the best in defense cost among all, followed by GANT and RNT. This is also a trend that is in line with that exhibited in Fig.~\ref{fig:base_ap} for deception effectiveness. DPNT has the lowest $C_D$ among all due to less edge changes made during topology shuffling compared to GANT and RNT. 
\item
Fig.~\ref{fig:base_ratio} compares packet delivery ratio PDR (the higher the better) representing service availability. In the ``when-to-shuffle'' category, adaptive/hybrid shuffling (AS/HS) based schemes perform comparably among themselves and outperform fixed/random shuffling (FS/RS) based schemes. The reason is that AS/HS produces a smaller number of attack paths toward decoy targets than FS/RS, so the attacker has a smaller chance to drop or manipulate packets passing through the attack paths. In the ``how-to-shuffle'' category, GANT, DPNT and RNT perform comparably among themselves. 
\end{itemize}

Summarizing above, there is no winner that can achieve the goal of maximizing deception effectiveness (see Fig.~\ref{fig:base_ap}), MTTSF (see Fig.~\ref{fig:base_mttsf}), and service availability (see Fig.~\ref{fig:base_ratio}) while minimizing defense cost (see Fig.~\ref{fig:base_cost}). However, we could identify DPNT as the best ``how-to-shuffle'' strategy that can maximize MTTSF and minimize defense cost, while maintaining comparable service availability. We explore optimal parameters of DPNT-based schemes in Section~\ref{sec:sensitivity} and compare the IoT network with and without these DPNT-based schemes in Section~\ref{sec:comparision-with-vs-without-NTS-MTD-running}. 
We also note that RS performs better than FS even the mean time interval for executing topology shuffling in RS is the same as the fixed time interval for executing topology shuffling in FS (see Table~\ref{tab:parameters}). We attribute this to the fact that the execution time interval in RS follows exponential distribution and this stochastic nature matches better with the stochastic nature of attack behavior. 


\subsubsection{Analysis on Impact of Attacker's Intelligence} \label{sssec:attack-intelligence}
We use the baseline scenario, except considering attackers with different levels of intelligence. We consider three levels of attack intelligence represented by three pairs of interaction probabilities with decoys ($P_d^{em}$ for an emulated decoy, $P_d^{os}$ for a full-OS based decoy): low intelligence (0.9, 1.0), medium intelligence (0.5, 0.7) and high intelligence (0.1, 0.3). For other design parameters, we follow their default values summarized in Table \ref{tab:parameters}. Without loss of generality, we consider AS-DPNT and HS-DPNT to analyze the impact of attack intelligence.

\begin{figure*}
  \centering
  \subfloat[$N_{DT}^{AP}$ ]{\includegraphics[width=0.5\textwidth]{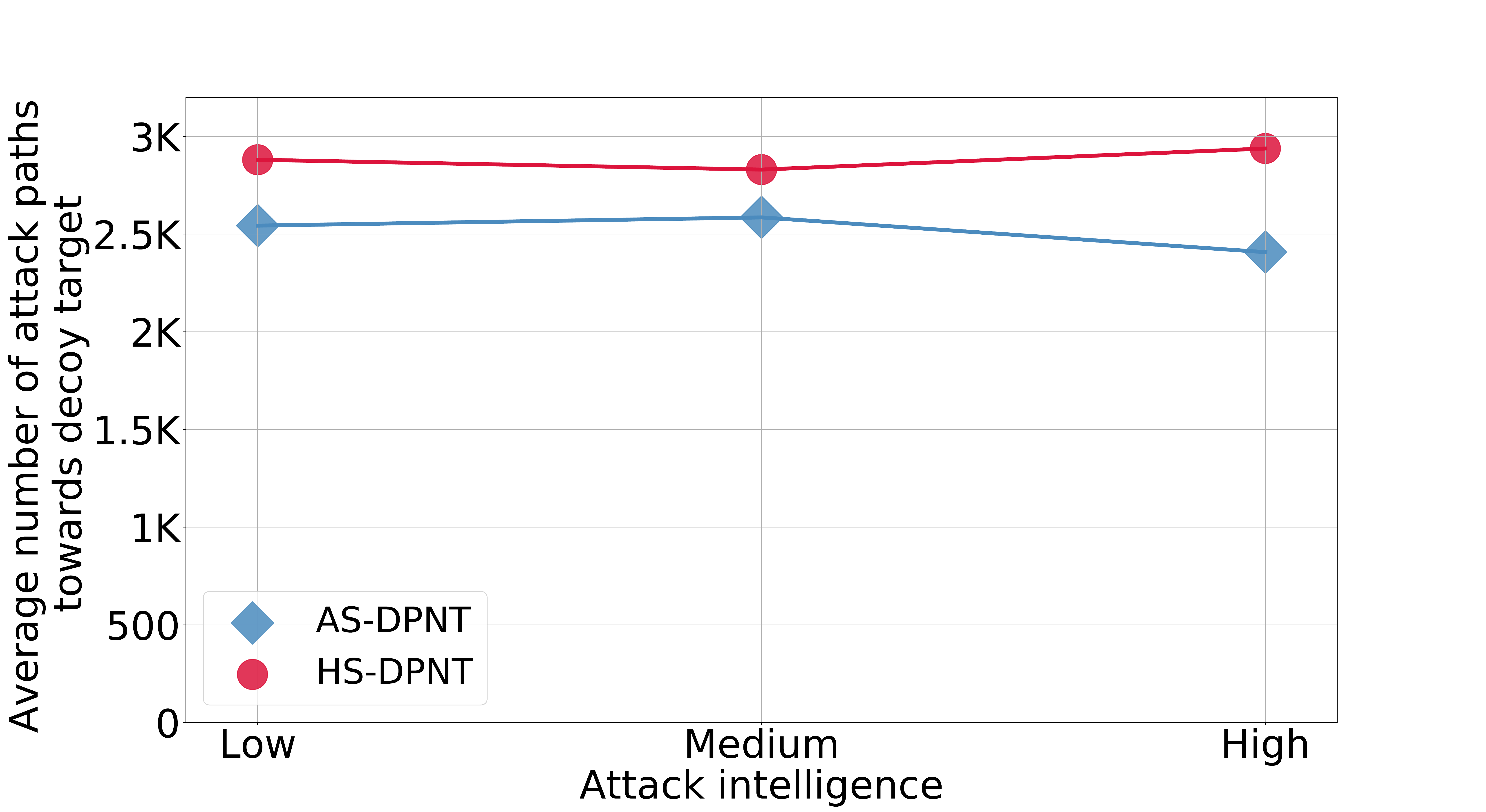}\label{fig:attack_ap}}
  \hfill
  \subfloat[MTTSF]{\includegraphics[width=0.5\textwidth]{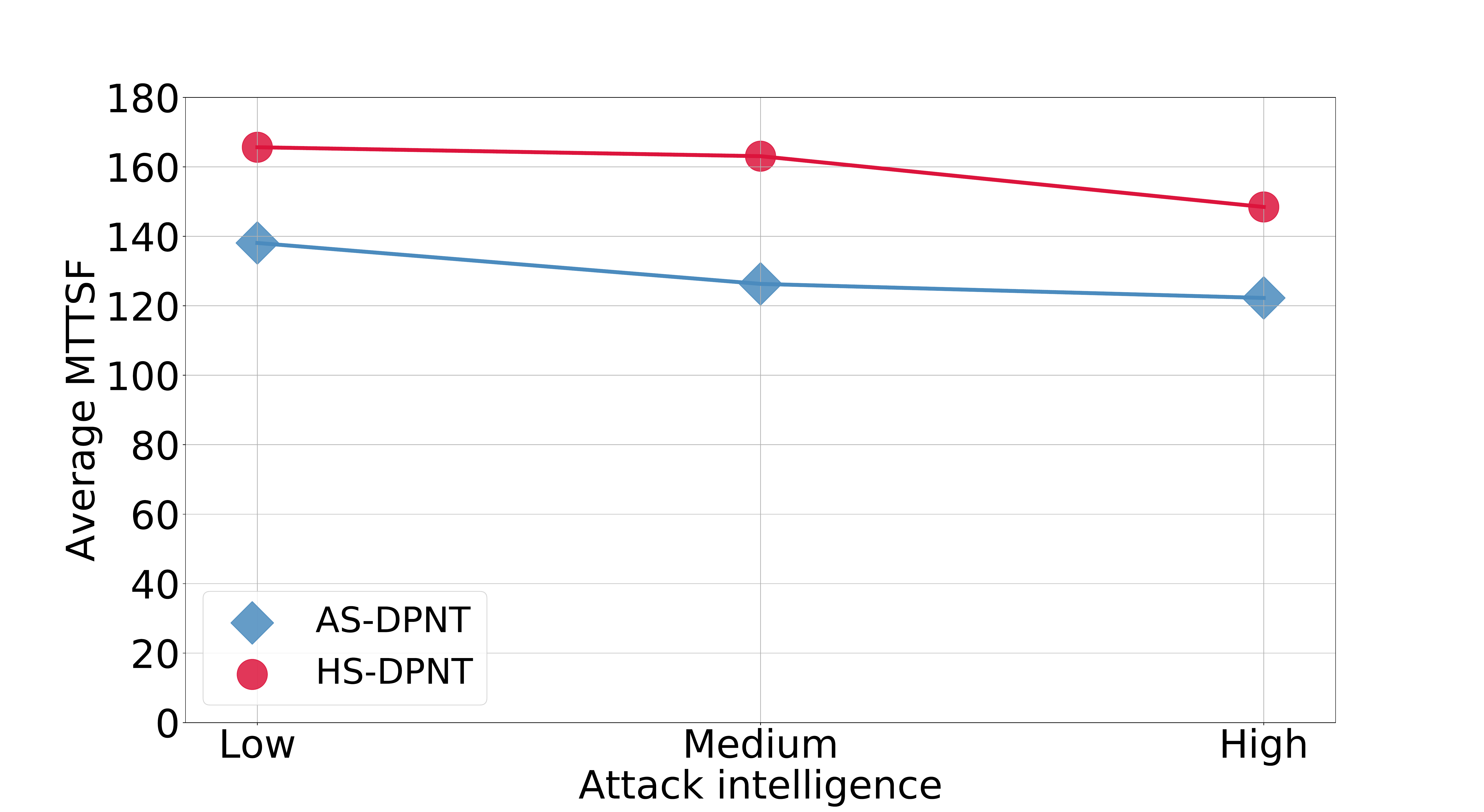}\label{fig:attack_mttsf}}
  \hfill
  \subfloat[$C_D$]{\includegraphics[width=0.5\textwidth]{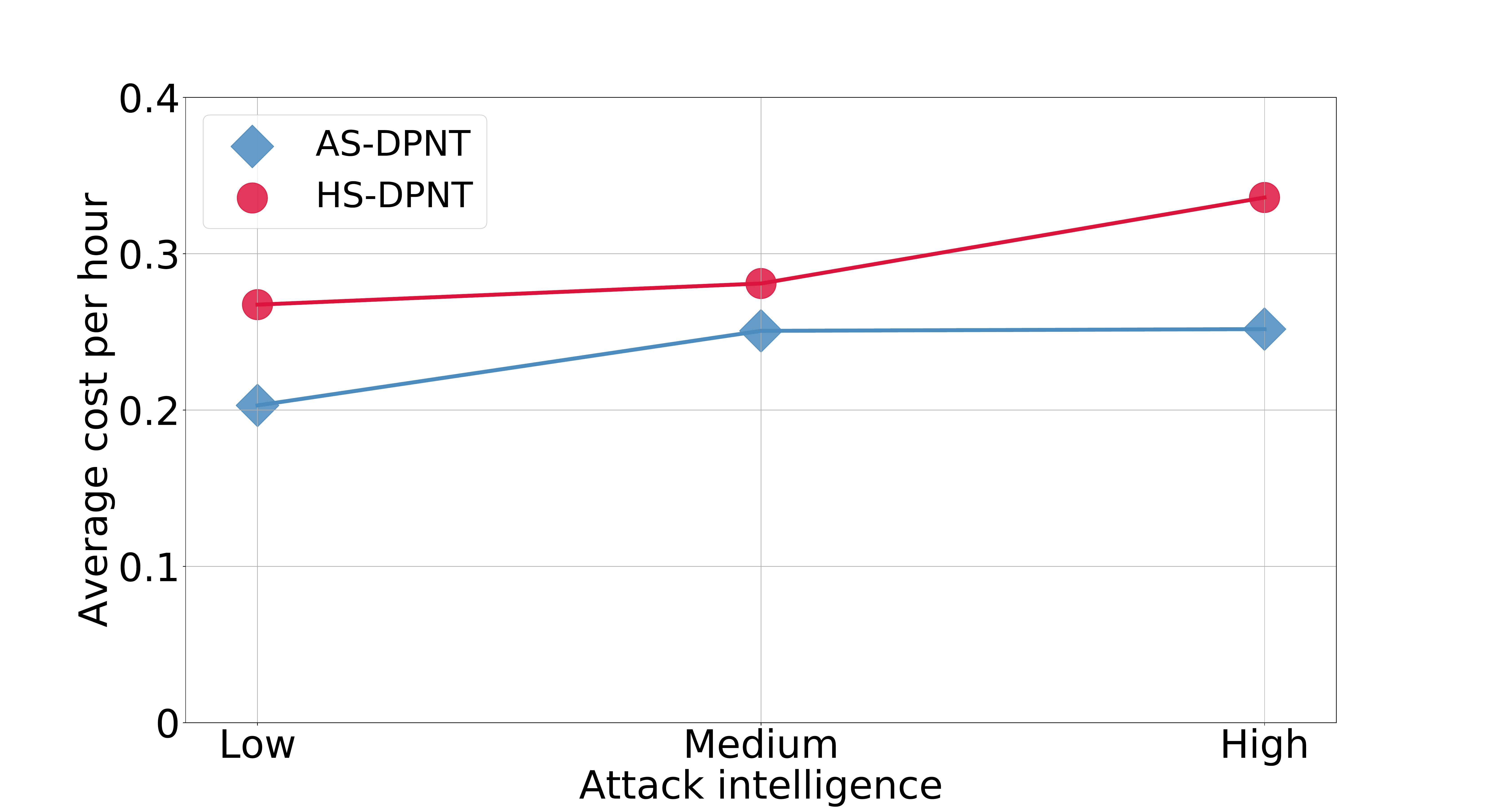}\label{fig:attack_cost}}
  \hfill
  \subfloat[PDR]{\includegraphics[width=0.5\textwidth]{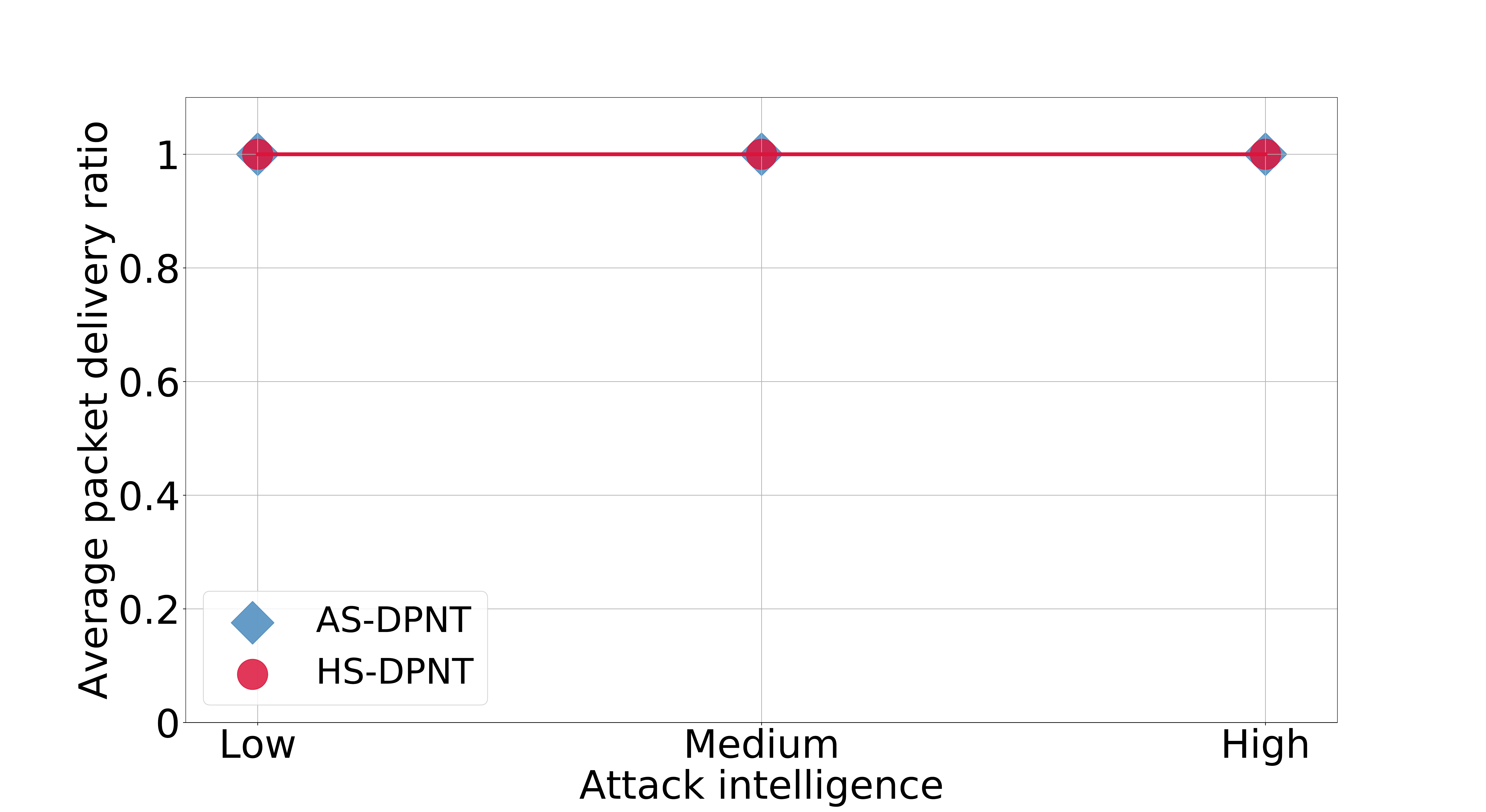}\label{fig:attack_ratio}}
  \caption{Performance analysis on impact of an attacker's intelligence.}\label{fig:attack-intelligence}
\end{figure*}

Fig.~\ref{fig:attack-intelligence} shows how AS-DPNT and HS-DPNT perform in terms of the mean time to security failure, MTTSF, 
the number of attack paths toward decoy targets, $N_{DT}^{AP}$, 
the defense cost per time unit, $C_D$, 
and the packet delivery ratio, PDR. In Fig.~\ref{fig:attack_ap}, with the decreasing attack intelligence, $N_{DT}^{AP}$ fluctuates for each scheme as this metric is related to the shuffling algorithm (i.e., DPNT in this case study).  In Fig.~\ref{fig:attack_mttsf}, for each scheme, MTTSF reaches the highest when the attacker has low intelligence. This implies the potential attacker with higher intelligence in detecting decoys hurts the system lifetime as measured based on MTTSF. However, both AS-DPNT and HS-DPNT are resilient under high-intelligent attacks without much reduction of MTTSF compared with the case of low-intelligent attacks. In Fig.~\ref{fig:attack_cost}, $C_D$ has an increasing trend for both schemes when intelligence increases. In Fig.~\ref{fig:attack_ratio}, PDR remains at 1.0 for both schemes. One reason is that in adaptive/hybrid shuffling, critical nodes may not be compromised when the $SSV$ threshold is small (e.g., $\rho$ = 0.1). Even if some neighbor nodes are compromised, there are still some clean neighbor nodes to be able to deliver packets. Another reason is that in our simulation setting, $P_a^d$ = 0.5 and $P_a^m$ = 0 to avoid detection, so compromised nodes will only drop half of the packets passing through them. 

In summary, attack intelligence has a moderate degree of impact (10-20\%) on MTTSF and $C_D$ because high intelligent attackers are capable of detecting decoys early on. This allows them to have more interactions with real nodes early on, thereby leading to shorter lifetime and forcing the system to trigger costly shuffling operations to prevent security attacks. Attack intelligence, however, has little impact on $N_{DT}^{AP}$ and PDR.

\subsubsection{Analysis on Impact of Attack Severity on Service Availability} \label{sssec:attack-severity}
We use the baseline scenario, except considering attacks with different levels of severity that would affect service availability. We consider three levels of attack severity represented by three pairs of packet drop probability $P_a^d$ and packet manipulation probability $P_a^m$: low severity (0.1, 0.1), medium severity (0.5, 0.5) and high severity (1.0, 1.0). For other design parameters, we follow their default values summarized in Table~\ref{tab:parameters}. We again apply DPNT-based schemes to analyze the impact of attack severity on packet delivery ratio (PDR) representing service availability.

\begin{figure}[htbp]
    \centering
    \includegraphics[width=0.5\textwidth]{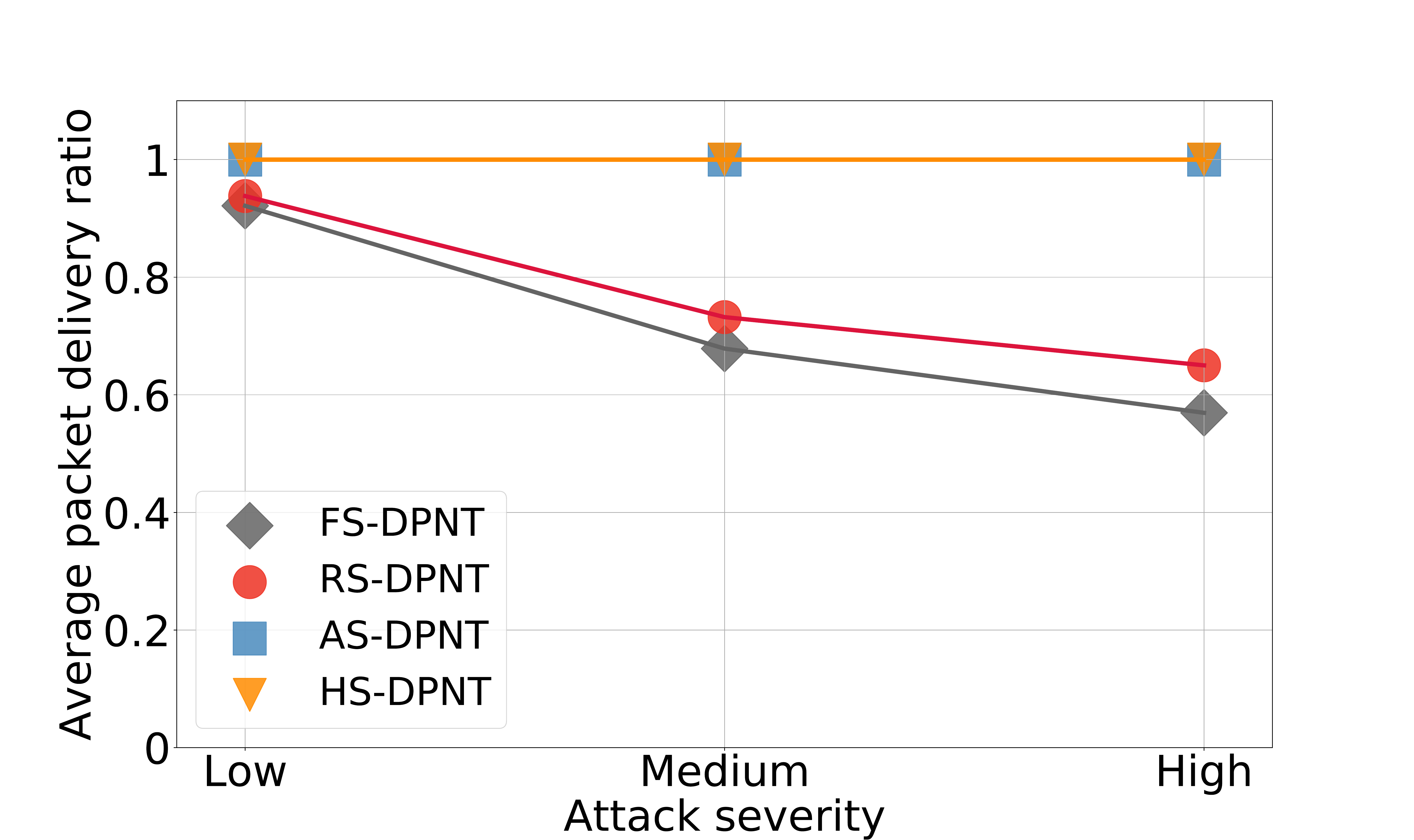}
\caption{Comparative performance analysis of the variants of DPNT schemes under the different attack severity.}
\label{fig:attack_severity}
\end{figure} 

Fig.~\ref{fig:attack_severity} shows the effect attack severity on PDR for DPNT based schemes. We observe that PDR remains at 1.0 for AS-DPNT and HS-DPNT while steadily decreases for FS-DPNT and RS-DPNT as the attack severity increases. This demonstrates resilience of adaptive shuffling schemes (i.e., AS/HS) in response to increasing attack severity because critical nodes are well protected from security attacks by setting a low $SSV$ threshold (e.g., $\rho$ = 0.1).

\subsubsection{Analysis on Impact of Decoy Node Population} \label{sssec:decoy}
We increase the number of decoy nodes in each VLAN to analyze the impact of decoy node population. The baseline scenario has (1, 1, 1, 1) decoy nodes for 
(MRI/CT scan, smart thermostat/meter/camera, smart TV/laptop, server).
We consider two more scenarios: (2, 2, 2, 2) and (3, 3, 3, 3). For other design parameters, we follow their default values summarized in Table~\ref{tab:parameters}. We consider AS-DPNT and HS-DPNT to analyze the impact of decoy population. 

\begin{figure*}
  \centering
  \subfloat[$N_{DT}^{AP}$ ]{\includegraphics[width=0.5\textwidth]{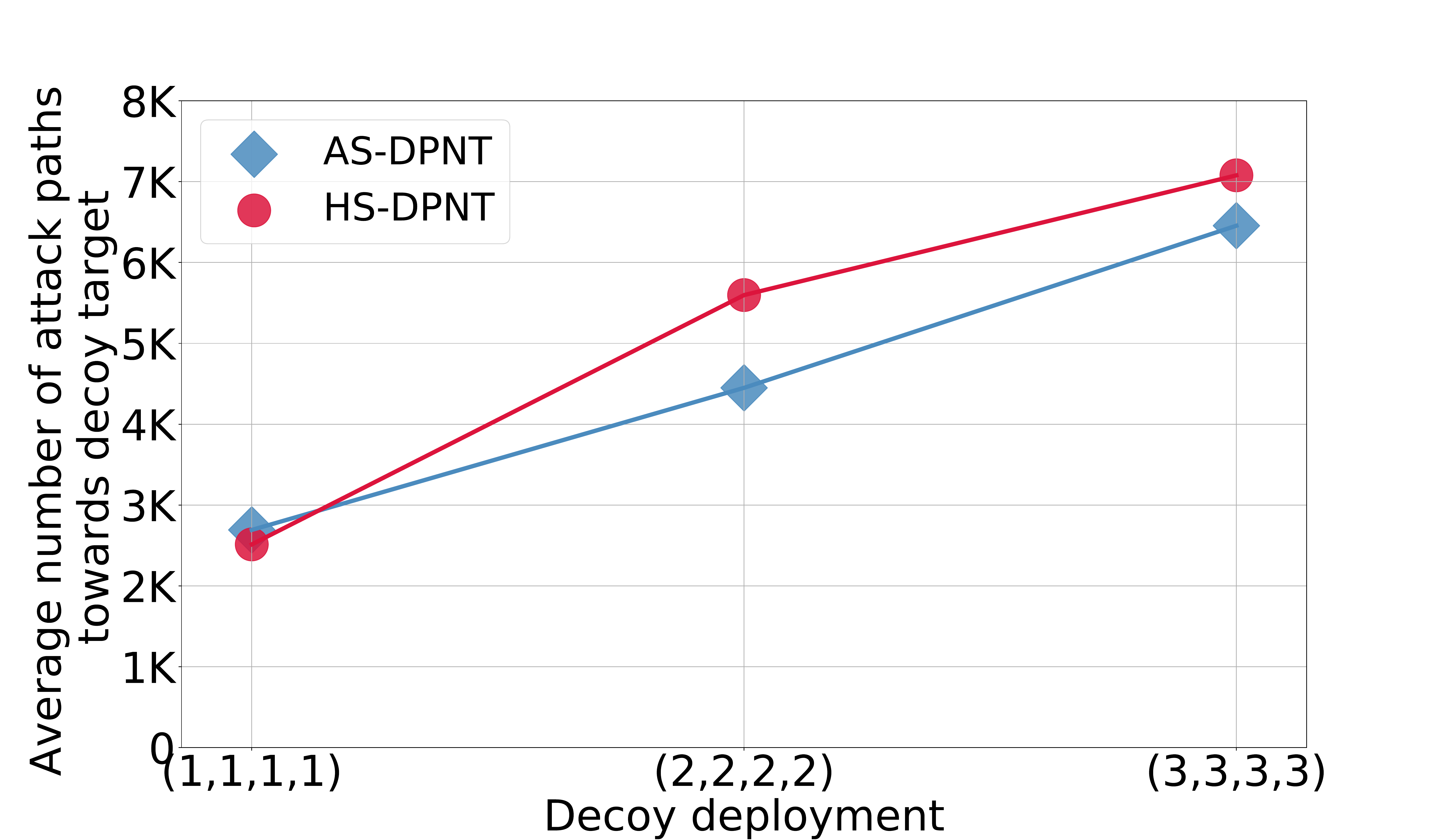}\label{fig:decoy_ap}}
  \hfill
  \subfloat[MTTSF]{\includegraphics[width=0.5\textwidth]{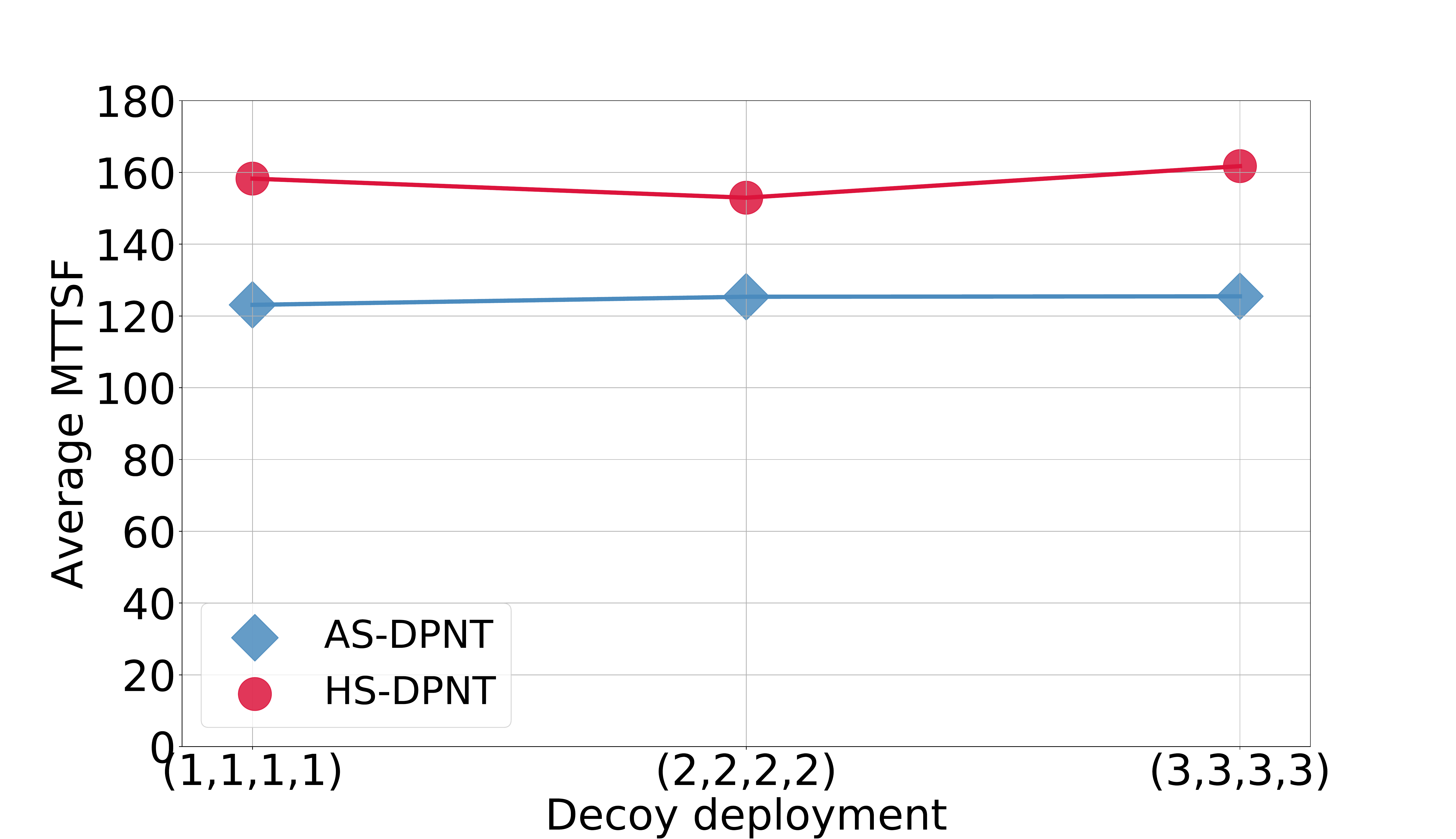}\label{fig:decoy_mttsf}}
  \hfill
  \subfloat[$C_D$]{\includegraphics[width=0.5\textwidth]{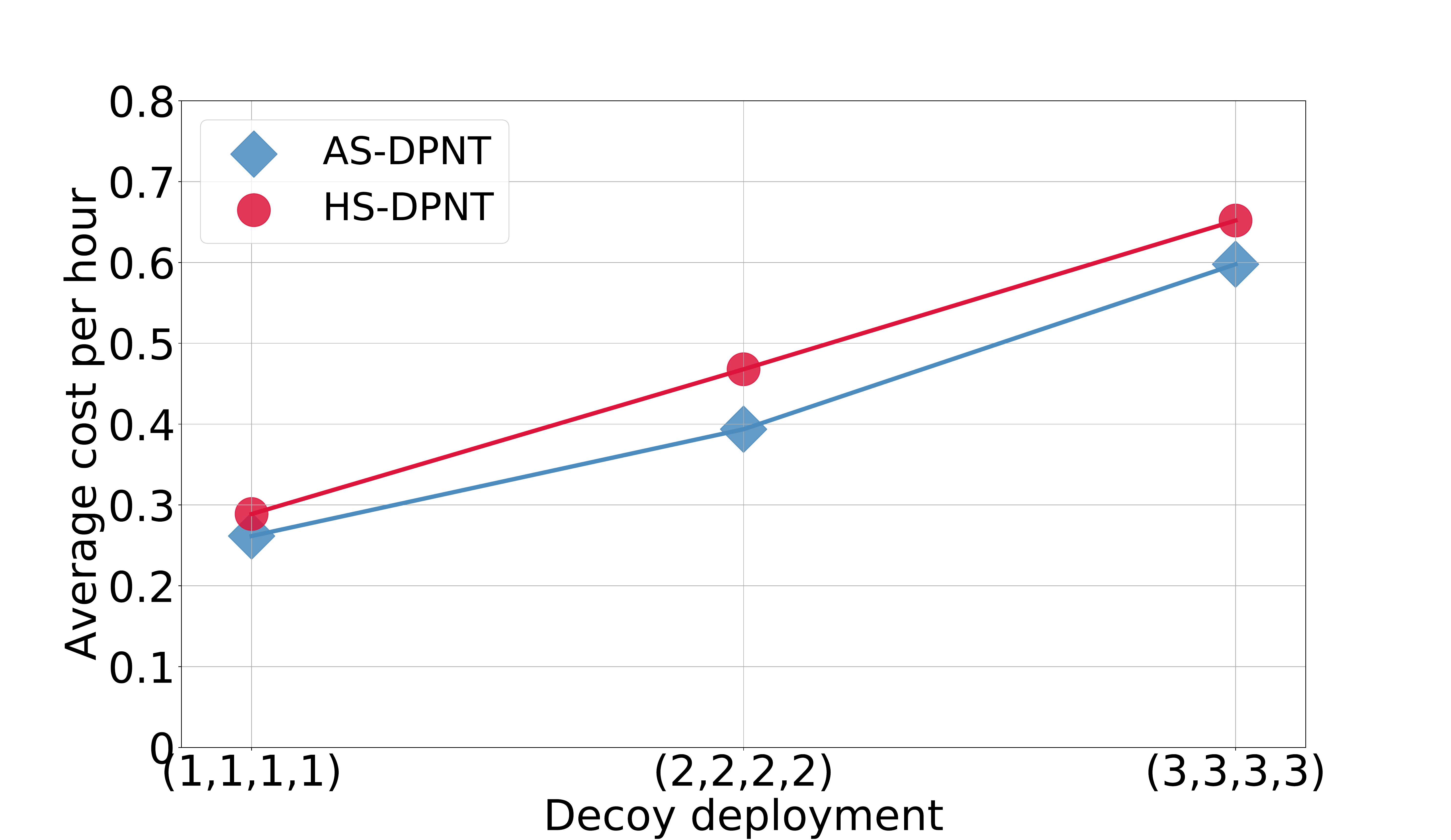}\label{fig:decoy_cost}}
  \hfill
  \subfloat[PDR]{\includegraphics[width=0.5\textwidth]{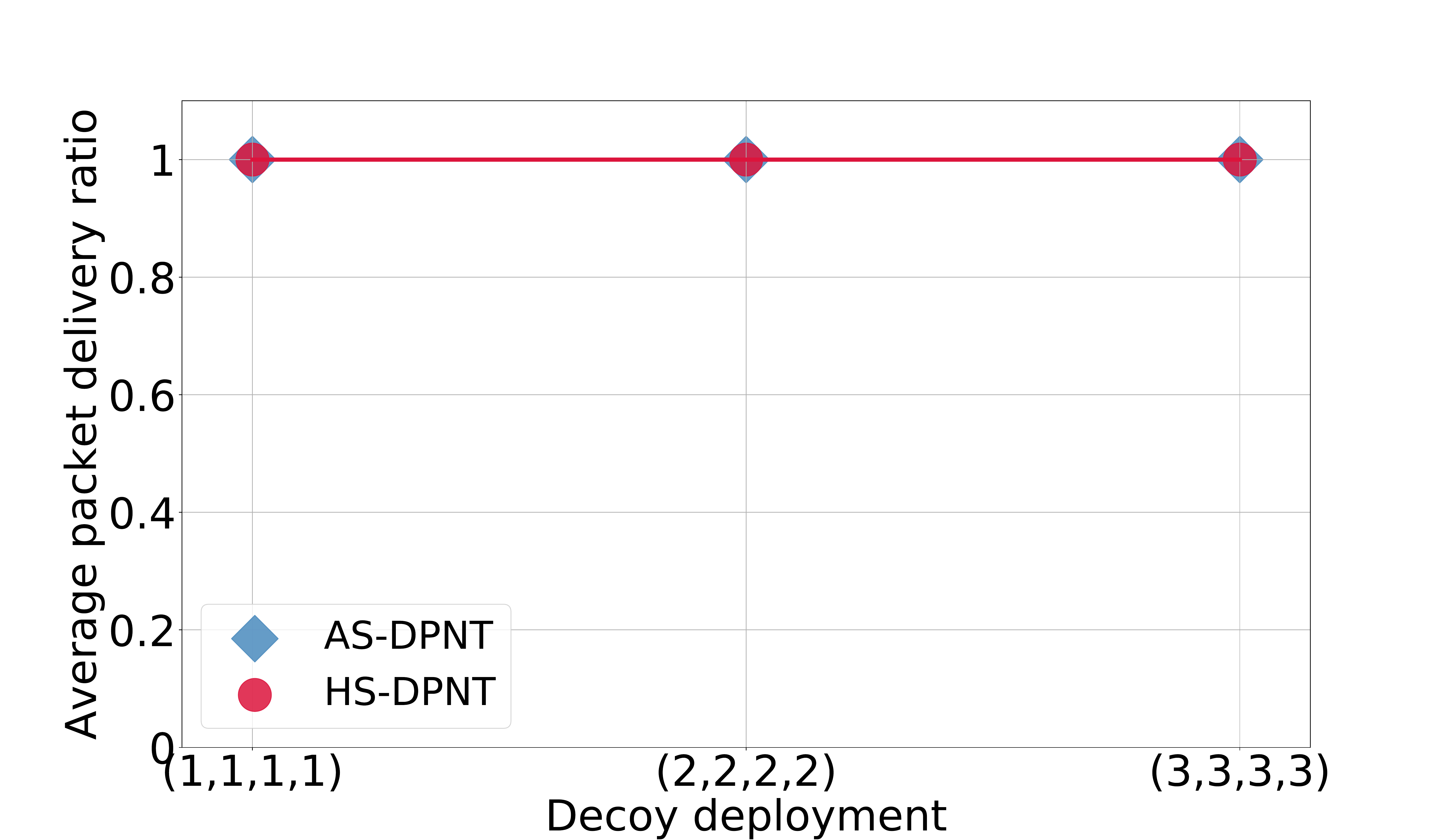}\label{fig:decoy_ratio}}
  \caption{Performance comparative analysis of the variants of DPNT schemes under different decoy deployment scenarios.}\label{fig:decoy}
\end{figure*}

Fig.~\ref{fig:decoy} shows how AS-DPNT and HS-DPNT perform in terms of the mean time to security failure, MTTSF, the number of attack paths toward decoy targets, $N_{DT}^{AP}$, the defense cost per time unit, $C_D$, and the packet delivery ratio, PDR, as the decoy population changes.

In Fig.~\ref{fig:decoy_ap}, $N_{DT}^{AP}$ increases significantly with the increasing number of decoys within each scheme. 
The reason is that as the number of decoys increases, DPNT also increases the number of attack paths toward decoy targets. In Fig.~\ref{fig:decoy_mttsf}, MTTSF remains steady as the number of decoys increases. We attribute this to the design of DPNT algorithm which only focuses on maximizing the number of decoy paths. There may be many paths with a majority of real IoT nodes on the paths. Attackers could still be able to compromise a large portion of real IoT nodes thus leading to system failure because MTTSF is calculated based on compromised real nodes within the network. 
In Fig.~\ref{fig:decoy_cost}, $C_D$ also increases as the number of decoys increases. This is due to the fact that a lot of edges need to be changed with additional decoy nodes.  In Fig.~\ref{fig:decoy_ratio}, PDR remains at 1.0 across all scenarios as PDR is related to service availability among real nodes, so it is little affected by decoys especially when $SSV$ threshold is low (e.g., $\rho$ = 0.1 in our test case). 

In summary, decoy node population has a great impact on $N_{DT}^{AP}$ and $C_D$ while it does not largely improve MTTSF or PDR. 

\subsubsection{Analysis on Impact of Network Size (Real Node Population)}
\label{sssec:network-size}
We increase the number of real nodes in each VLAN to analyze the impact of network size (or real node population). 
The baseline scenario has (2, 3, 2, 1) real IoT nodes for (MRI/CT scan, smart thermostat/meter/camera, smart TV/laptop, server).
We consider one more scenario: (2, 6, 4, 1) real IoT nodes. 
The number of medical devices in VLAN1 is kept the same as in the baseline scenario as they are rarely deployed in a large scale due to their high price. Therefore, we have two scenarios with the number of real IoT nodes (excluding the server) as 7 and 12 respectively. We apply HS-DPNT to analyze the impact of network size.

\begin{table*}[h]
\renewcommand{\arraystretch}{1.3}
\caption{Analysis on impact of network size.}
    \centering
    \begin{tabular}{|c|c|c|}
    \hline
    \multirow{2}{*}{Metric}  & \multicolumn{2}{c|}{No. of real IoT nodes} \\
    \cline{2-3}
    & 7 & 12\\
    \hline
     $N_{DT}^{AP}$  & 2734.5 &  6206452.0 \\
    \hline
     MTTSF & 160.1 & 133.0 \\
    \hline
    $C_D$ & 0.23 & 0.53 \\
    \hline
    PDR & 1.0 & 1.0 \\
    \hline
    \end{tabular}
    \label{tab:network_size}
\end{table*}

Table~\ref{tab:network_size} shows the effect of network size on performance in terms of the mean time to security failure, MTTSF, the number of attack paths toward decoy targets, $N_{DT}^{AP}$, the defense cost per time unit, $C_D$, and the packet delivery ratio, PDR. We see that $N_{DT}^{AP}$ has a significant jump from 2734.5 to 6206452.0. As the number of real IoT nodes increases, the number of decoy paths with real IoT nodes acting as entry points and intermediate nodes also increases. Since more decoy paths/edges are created as the number of real IoT nodes increases, the defense cost $C_D$ doubles.
MTTSF decreases by 16.9\% as the number of real IoT nodes increases. The reason is that more real nodes introduce more attack surface and as more real nodes are compromised the system failure condition {\tt SFC1} specified in Section~\ref{sec:security-fail-conditions} can trigger a failure. Lastly network size has little effect on PDR due to the same reason explained earlier in Section~\ref{sssec:attack-intelligence}. 

In summary, network size (i.e., real node population) has a high impact on $N_{DT}^{AP}$, $C_D$, and MTTSF, but little impact on PDR. 

\subsection{Sensitivity Analysis}\label{sec:sensitivity}
In this section, we examine the sensitivity of the performance results
with respect to the maximum delay parameter ($\gamma_2$) and the security vulnerability level (SSV) threshold parameter ($\rho$) to identify the optimal parameter setting under which the system performance can be maximized. These two parameters are used in two ``when-to-shuffle'' strategies, namely, adaptive shuffling and hybrid shuffling (AS/HS). Without loss of generality, we consider HS-DPNT in the sensitivity analysis since earlier we have identified DPNT as the best ``how-to-shuffle'' strategy that can maximize MTTSF and minimize defense cost, while maintaining comparable service availability.

\subsubsection{Sensitivity Analysis of Maximum Delay}
We use the baseline scenario in Section \ref{sssec:compare-scheme}, except that we vary the maximum delay parameter, $\gamma_2$, when performing hybrid shuffling. The reason we use a maximum delay is to avoid the situation in which an incremental increase of $SSV$ does not reach the threshold, thus delaying the execution of DPNT. We consider the following values for $\gamma_2$ in the sensitivity analysis: 48, 72, 96, 120, 144, 168 (hours). These values are related to the scenario and could change due to different scenarios. For other design parameters, we follow their default values summarized in Table~\ref{tab:parameters}.

\begin{figure}
  \centering
  \subfloat[$N_{DT}^{AP}$ ]{\includegraphics[width=0.5\textwidth]{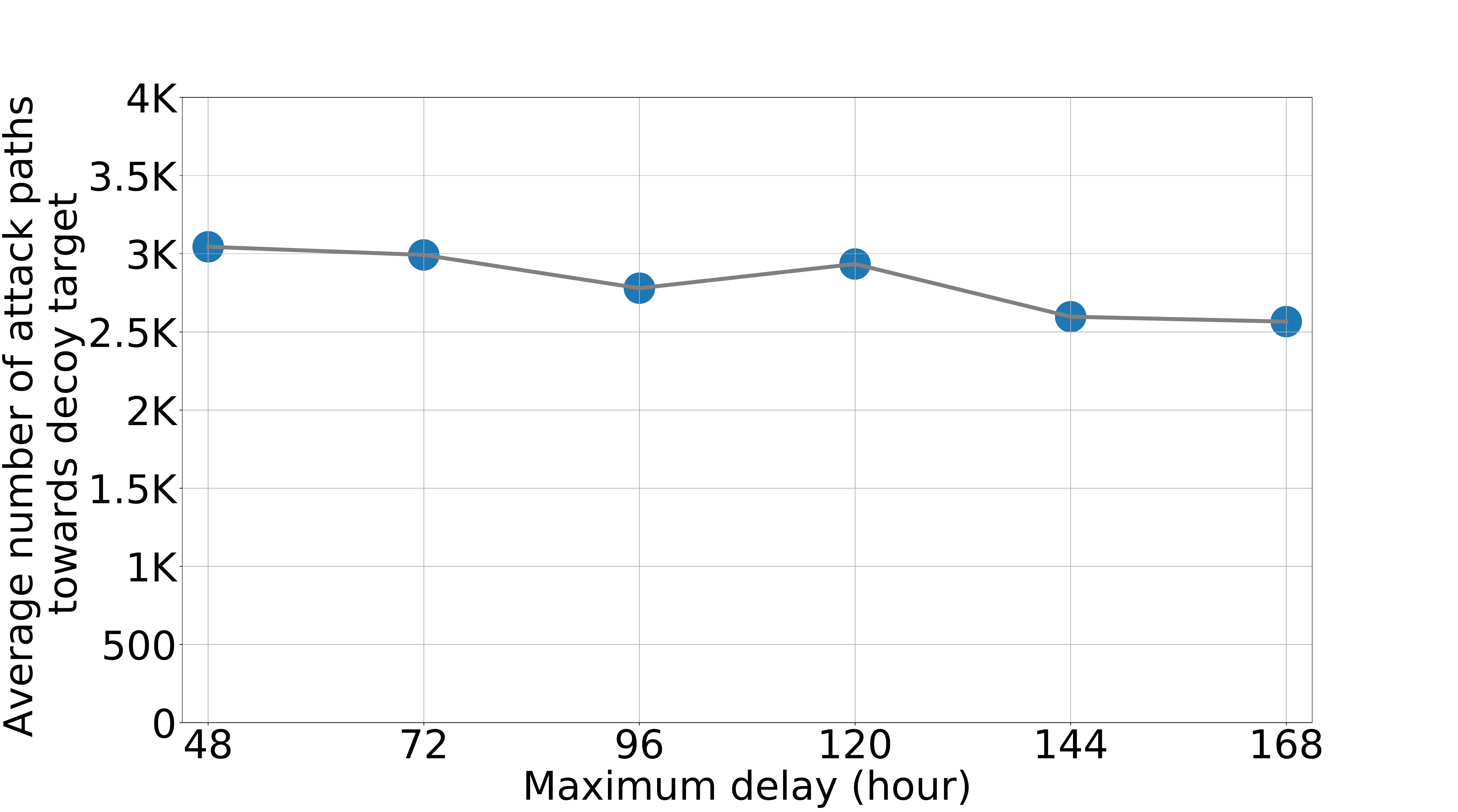}\label{fig:delay_ap}}
  \hfill
  \subfloat[MTTSF]{\includegraphics[width=0.5\textwidth]{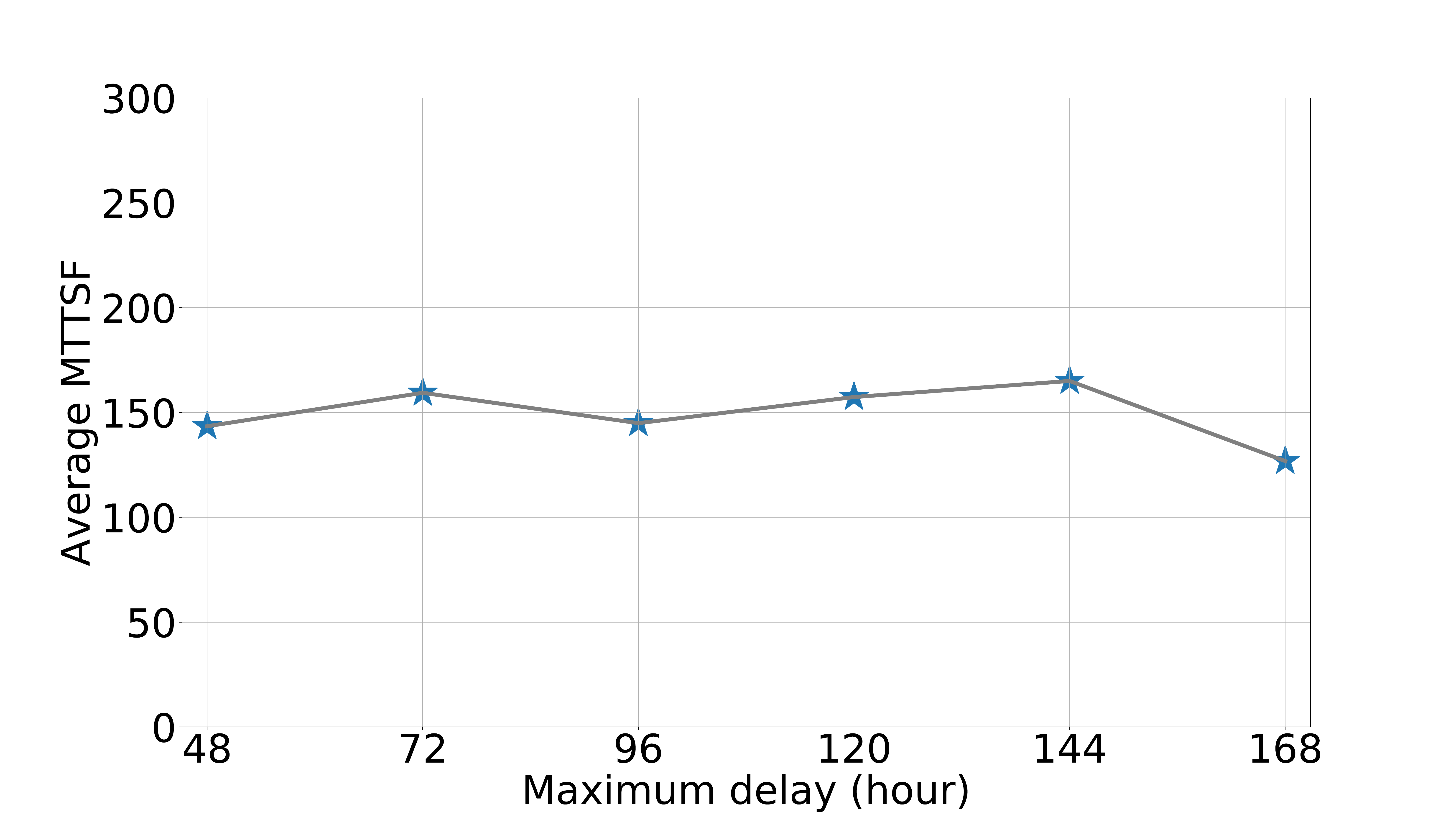}\label{fig:delay_mttsf}}
  \hfill
  \subfloat[$C_D$]{\includegraphics[width=0.5\textwidth]{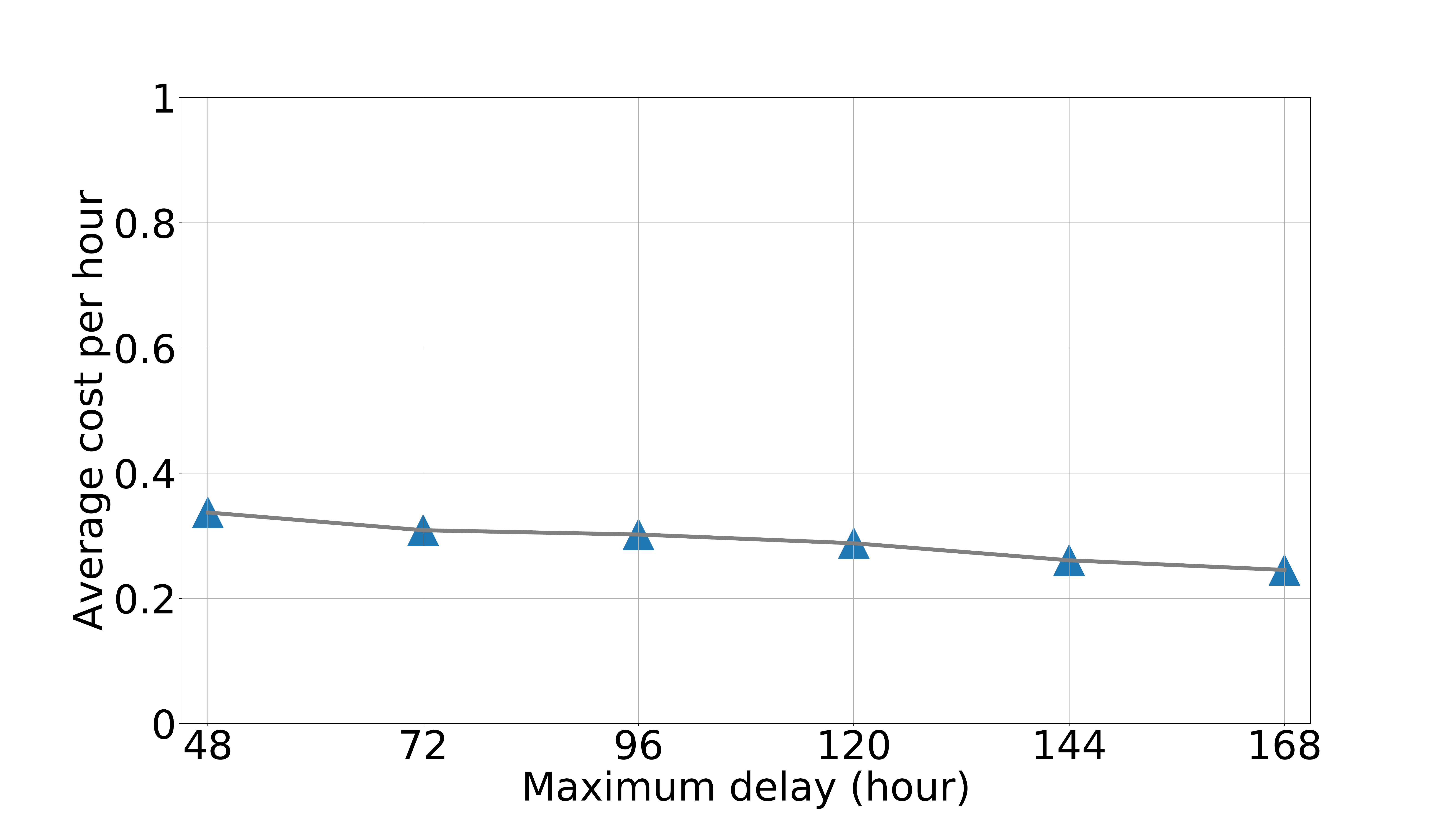}\label{fig:delay_cost}}
  \hfill
  \subfloat[PDR]{\includegraphics[width=0.5\textwidth]{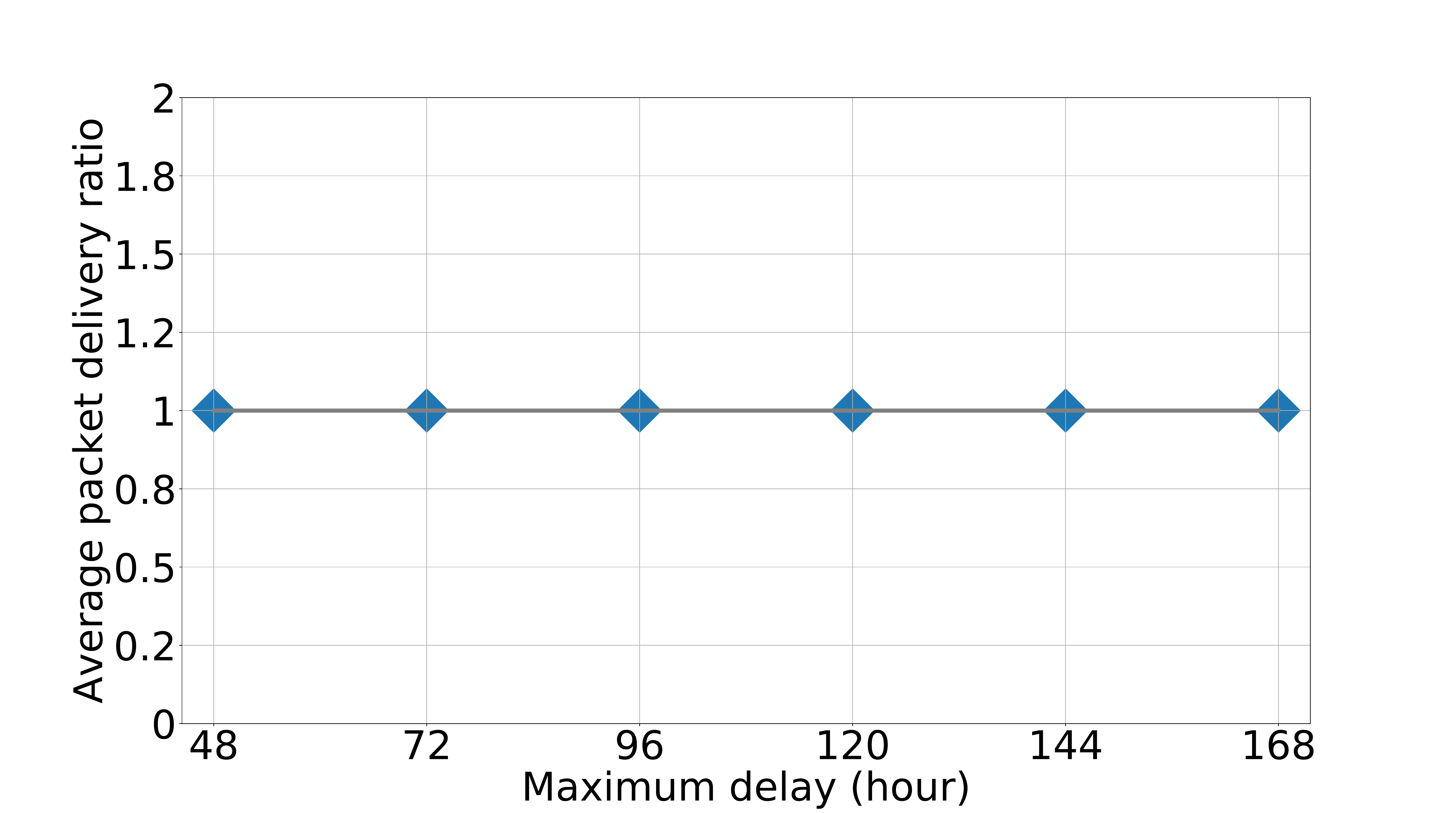}\label{fig:delay_ratio}}
  \caption{Effect of the maximum delay ($\gamma_2$) on the performance of HS-DPNT (identified as the best scheme) under the baseline scenario.}\label{fig:delay}
\end{figure}

Fig.~\ref{fig:delay} shows the sensitivity of the performance results in terms of $N_{DT}^{AP}$, MTTSF, $C_D$, and PDR with respect to the maximum delay parameter ($\gamma_2$) in hybrid shuffling (HS). Intuitively, a shorter delay may cause the network to be shuffled more often which makes HS similar to fixed/random shuffling (FS/RS) while a longer delay may delay shuffling thus degenerating HS to adaptive shuffling (AS). In Fig~\ref{fig:delay_ap}, $N_{DT}^{AP}$ fluctuates as the metric is related to how-to-shuffle instead of when-to-shuffle.  In Fig.~\ref{fig:delay_mttsf}, MTTSF fluctuates slightly from 48 to 120 with a local optimal value of 159 hours at 72, reaches the peak of 165 hours at 144 and then drops to 127 hours at 168.  In Fig.~\ref{fig:delay_cost}, $C_D$ steadily decreases as the maximum delay increases because topology shuffling will occur less frequently as the maximum delay increases. In Fig.~\ref{fig:delay_ratio}, PDR remains at 1.0. The reason is the same as stated in Section~\ref{sssec:attack-intelligence}. Summarizing above, if the goal is to maximize MTTSF, setting the maximum delay at 144 could be considered as optimal for HS-DPNT.

\subsubsection{Sensitivity Analysis of the System Security Vulnerability (SSV) Threshold}\label{secc:sensitivity-ssv}

We use the baseline scenario in Section~\ref{sssec:compare-scheme}, except that we vary the SSV threshold values, $\rho$, in the range of [0.1, 0.9] with 0.1 as the increment. For other design parameters, we follow their default values summarized in Table~\ref{tab:parameters}.  We again apply HS-DPNT in our sensitivity analysis. 

\begin{figure}
  \centering
  \subfloat[$N_{DT}^{AP}$ ]{\includegraphics[width=0.5\textwidth]{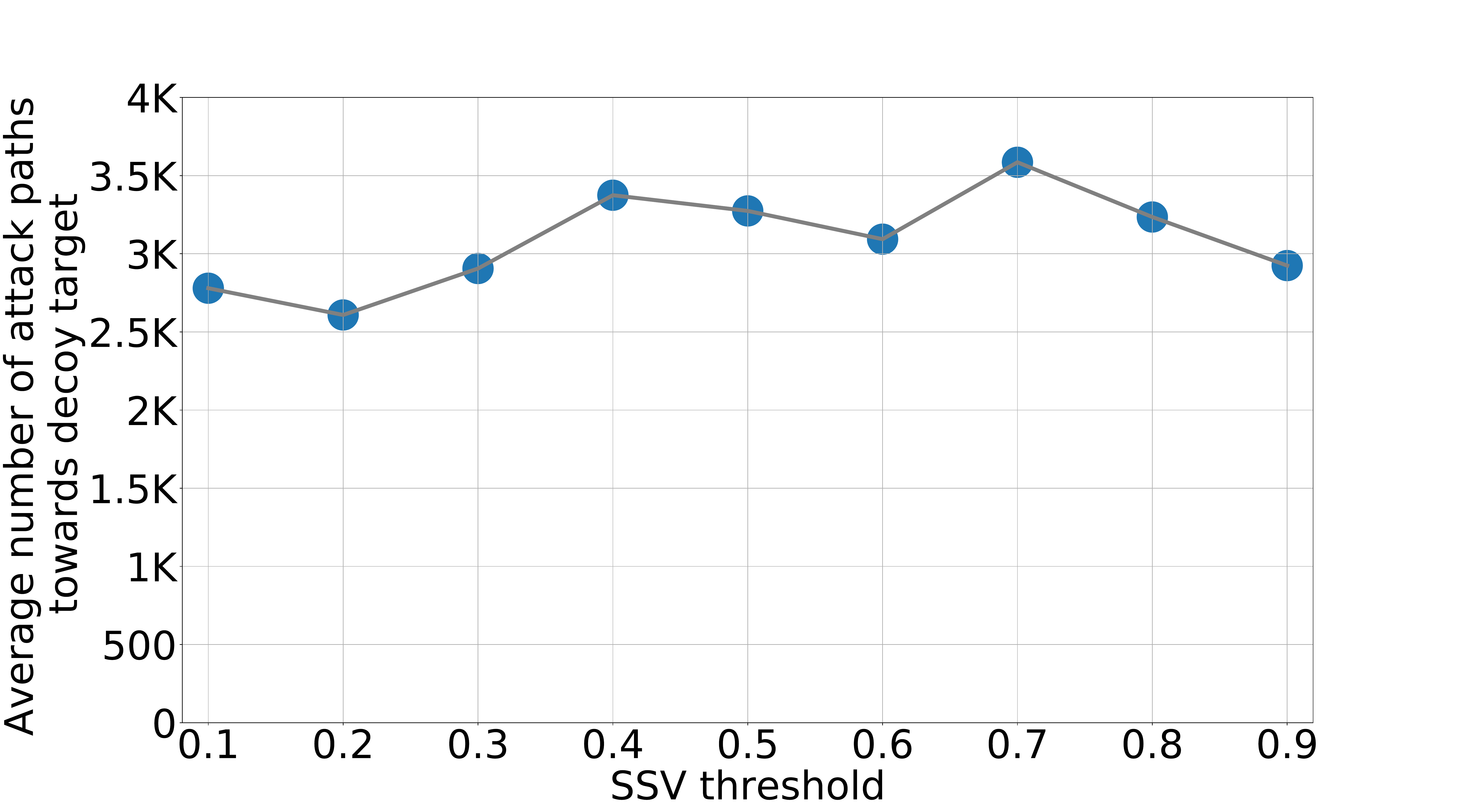}\label{fig:ssv_ap}}
  \hfill
  \subfloat[MTTSF]{\includegraphics[width=0.5\textwidth]{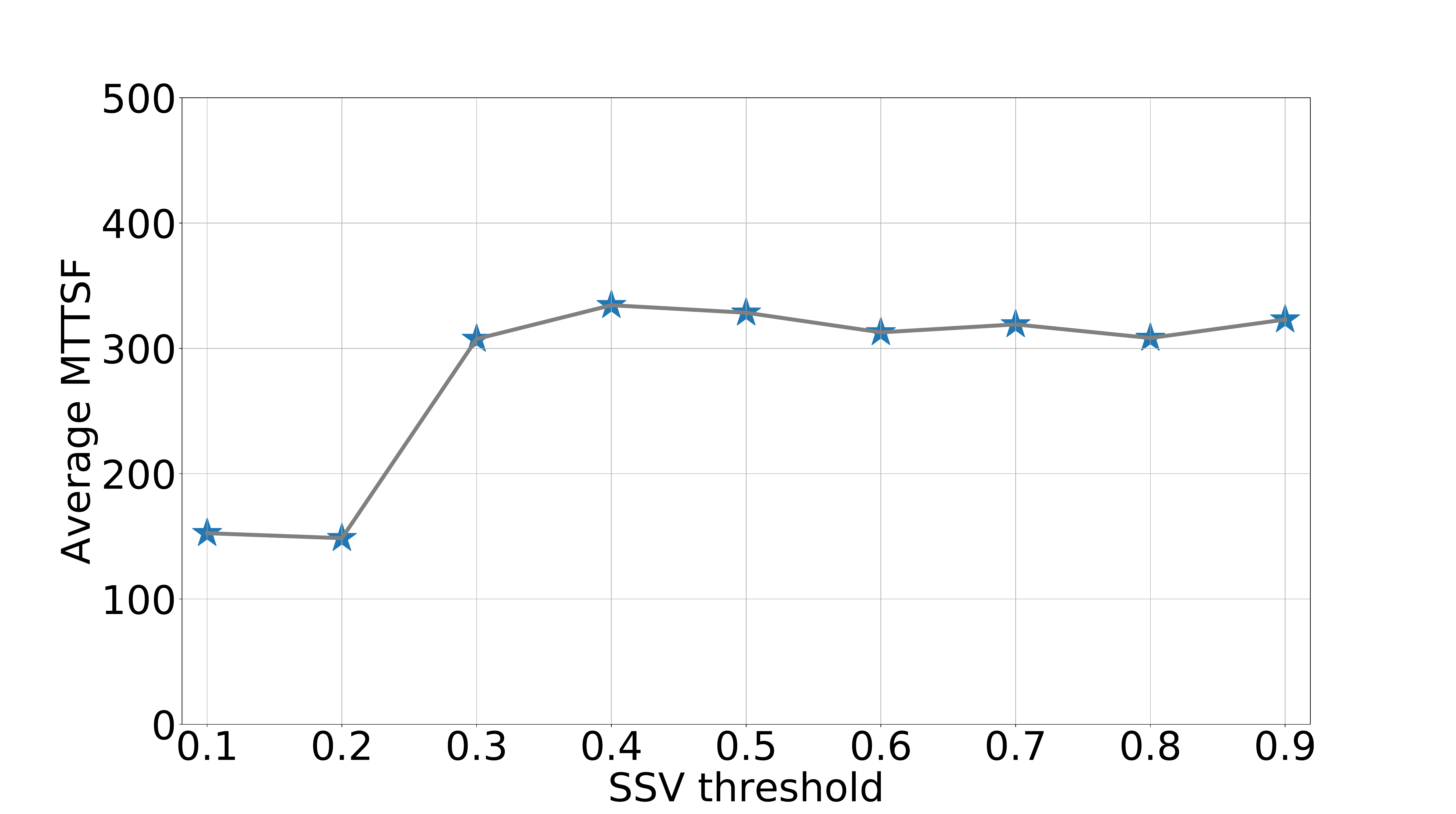}\label{fig:ssv_mttsf}}
  \hfill
  \subfloat[$C_D$]{\includegraphics[width=0.5\textwidth]{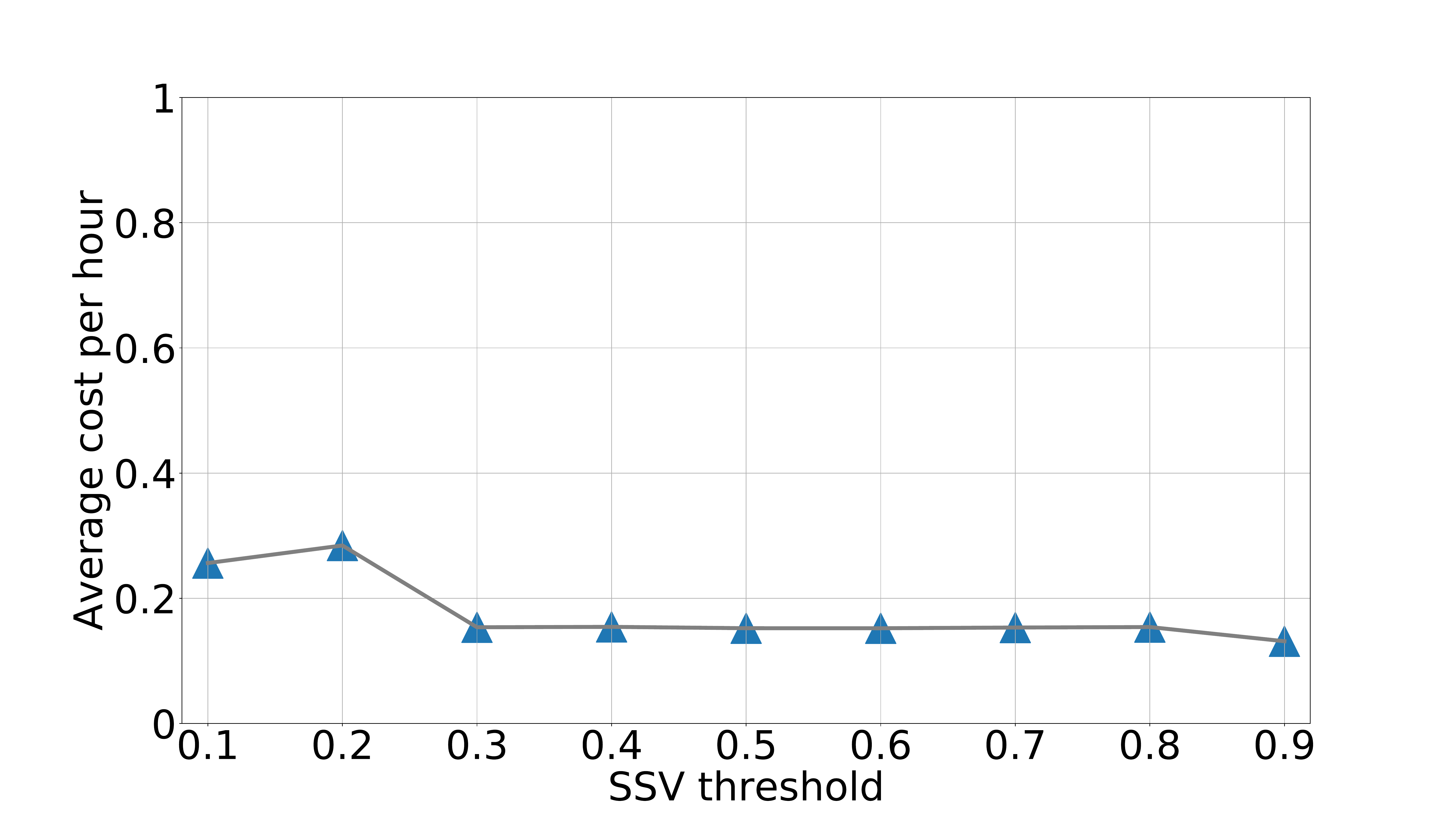}\label{fig:ssv_cost}}
  \hfill
  \subfloat[PDR]{\includegraphics[width=0.5\textwidth]{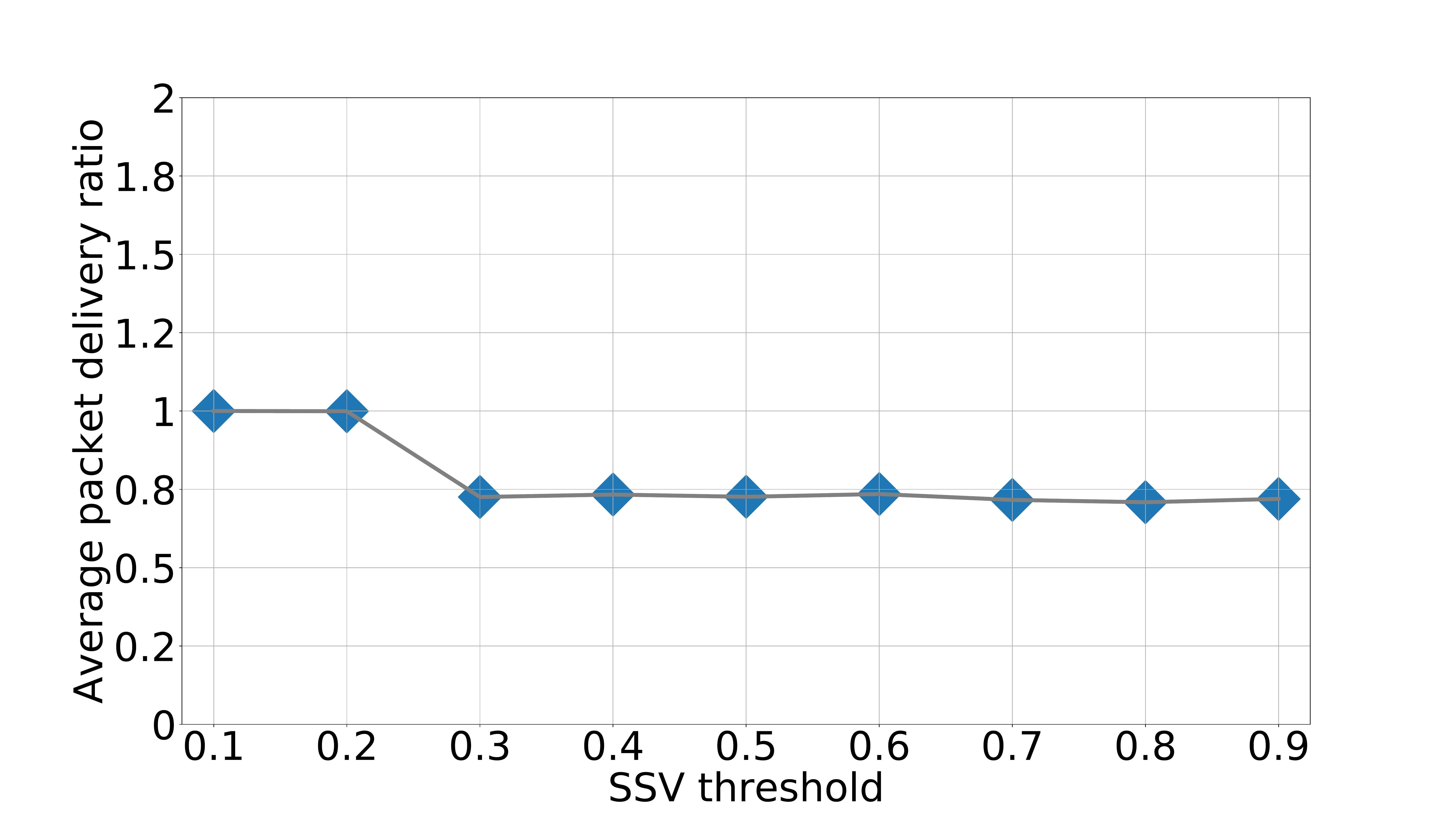}\label{fig:ssv_ratio}}
  \caption{Effect of the system security vulnerability ($SSV$) threshold $\rho$ on the performance of HS-DPNT (identified as the best scheme) under the baseline scenario.}\label{fig:ssv}
\end{figure}

Fig.~\ref{fig:ssv} shows the sensitivity of the performance results in terms of $N_{DT}^{AP}$, MTTSF, $C_D$, and PDR with respect to the SSV threshold parameter $\rho$. In Fig.~\ref{fig:ssv_ap}, $N_{DT}^{AP}$ fluctuates as $N_{DT}^{AP}$ is related to how-to-shuffle instead of when-to-shuffle. In Fig.~\ref{fig:ssv_mttsf}, MTTSF jumps from 149 at 0.2 to 308 at 0.3, slightly increases to 334 as the peak when $\rho$ is 0.4, and then varies between 308 and 328 when $\rho$ increases. 
In Fig.~\ref{fig:ssv_cost}, $C_D$ decreases to 0.15 when $\rho$ increases to 0.3 and then stays stable with increasing $\rho$. In Fig.~\ref{fig:ssv_ratio}, PDR drops to 0.73 when $\rho$ is 0.3 and stays stable afterwards. The reason is the same as stated in Section~\ref{sssec:attack-intelligence}. Summarizing above, if the goal is to maximize MTTSF, setting the SSV threshold parameter at 0.4 could be considered as optimal for HS-DPNT.

\subsection{Performance Comparison of IoT Networks with vs. without NTS-MTD Running} \label{sec:comparision-with-vs-without-NTS-MTD-running}
\begin{figure}
  \centering
  \subfloat[MTTSF]{\includegraphics[width=0.5\textwidth]{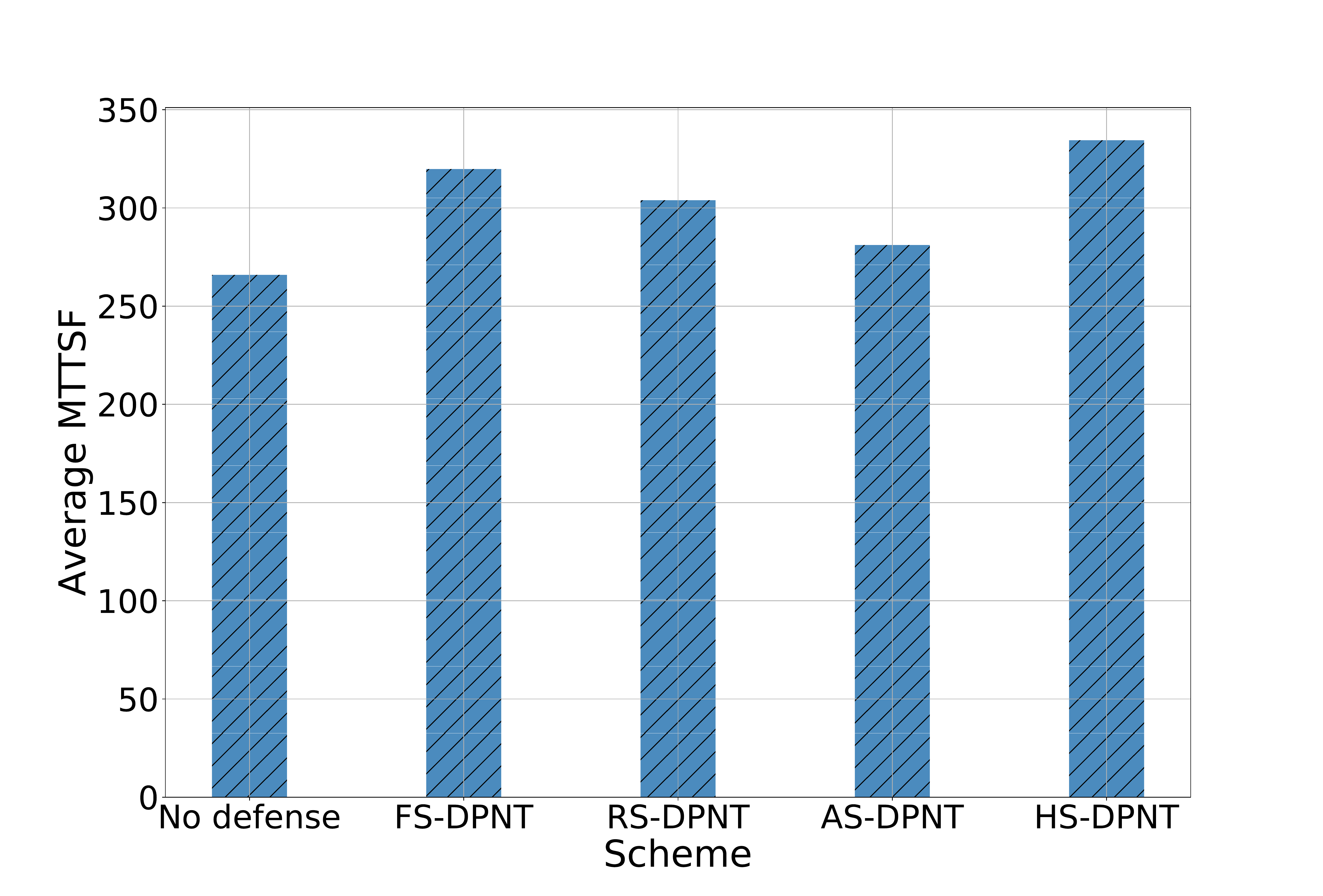}\label{fig:defense_mttsf}}
  \hfill
  \subfloat[PDR]{\includegraphics[width=0.5\textwidth]{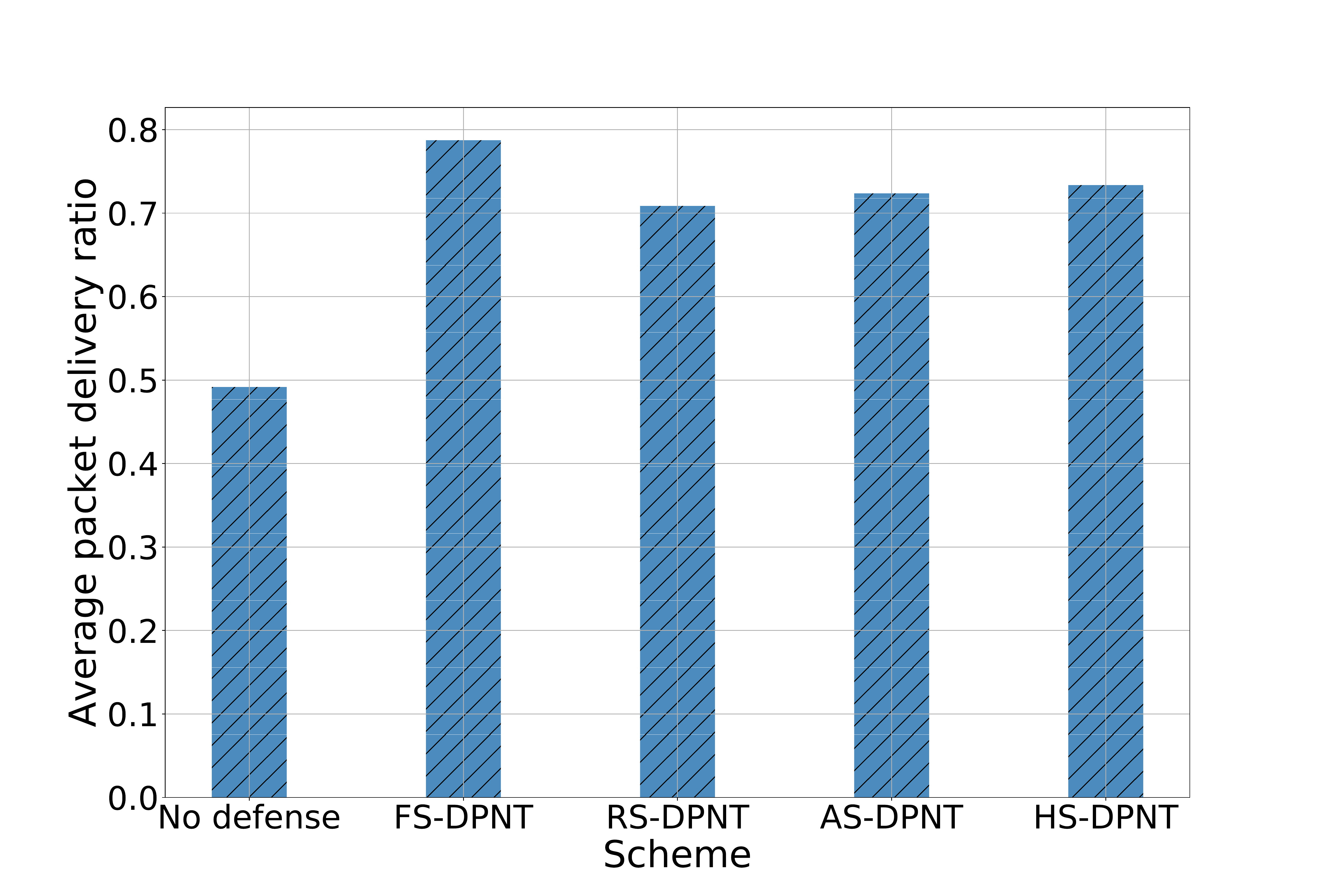}\label{fig:defense_pdr}}
  \caption{Performance Comparison of IoT Networks with vs. without NTS-MTD Running.}\label{fig:no_defense}
\end{figure}
In this section, we compare the performance of IoT networks with vs. without our proposed network topology shuffling-based MTD (NTS-MTD) technique running. 
That is, the baseline IoT network does not have decoy nodes deployed and does not have our proposed network topology shuffling-based MTD (NTS-MTD) technique running for intrusion prevention. We use the same baseline scenario as before with the same attack model applied. We collect performance data for computing MTTSF and PDR based on 100 times of simulation runs.  We compare the baseline scheme with DPNT based schemes running at optimal settings identified in the sensitivity analysis study of Section~\ref{sec:sensitivity} (i.e., FS-DPNT with the optimal fixed interval at 72 hours, RS-DPNT with the optimal mean interval at 24 hours, AS-DPNT with the optimal SSV threshold at 0.6, and HS-DPNT with the optimal SSV threshold at 0.4). Fig.~\ref{fig:no_defense} shows the performance comparison results 
in MTTSF and PDR over all DPNT based schemes considered in this work. The baseline IoT system is labeled with ``No defense'' in Fig.~\ref{fig:no_defense}. We observe that all DPNT-based schemes significantly outperform the counterpart baseline IoT system. In particular, HS-DPNT has the highest increase (26\%) in MTTSF while FS-DPNT has the highest increase (60\%) in PDR. These results demonstrate that an IoT network running our intrusion prevention technique at the optimal parameter setting prolongs system lifetime, increases attack complexity of compromising critical nodes (so that the system lifetime is prolonged), and maintains superior service availability compared with a counterpart baseline IoT network without running our intrusion prevention technique.

\section{Conclusions \& Future Work} \label{sec:conclusions}

In this paper, we proposed an integrated proactive defense mechanism by utilizing cyberdeception and network topology shuffling and completed a comprehensive analysis via 
simulation.
We considered a smart hospital scenario within the IoT context. The proposed approach could be applied to any IoT environment. 

From this study, we obtained the following {\bf key findings}: 
\begin{itemize}
\item In the ``when-to-shuffle'' category, adaptive/hybrid shuffling (AS/HS) based schemes outperform fixed/random shuffling (FS/RS) based schemes in defense cost. On the contrary, 
FS/RS based schemes outperform AS/HS based schemes in the average number of attack paths toward decoy targets (i.e., deception effectiveness). Choices of fixed/mean interval used by FS/RS and SSV threshold used by AS/HS have significant impact on MTTSF and service availability and need to be properly determined.
The analysis performed in this paper can help the system designer determine the best interval by FS/RS and best SSV threshold by AS/HS to maximize MTTSF. 
\item In the ``how-to-shuffle'' category, decoy path-optimized network topology (DPNT) based schemes perform comparably with genetic algorithm network topology (GANT) based schemes in MTTSF (i.e., system lifetime) and packet delivery ratio. On the other hand, DPNT incurs less defense cost than GANT since GANT tends to create more attack paths toward decoy targets (i.e., deception effectiveness). Both DPNT and GANT based schemes outperform random network topology (RNT) shuffling schemes in MTTSF and the number of attack paths toward decoy targets (deception efficiency). Consequently, if MTTSF is the goal, DPNT should be chosen over GANT because it incurs less defense cost while achieving comparable MTTSF. 
\item If maximizing MTTSF is the most important goal, while maximizing deception effectiveness and service availability and minimizing defense cost are sub-goals, HS-DPNT with an optimal SSV threshold (with HS as the ``when-to-shuffle'' strategy and DPNT as the ``how-to-shuffle'' strategy) emerges as the best scheme among the 12 schemes investigated for executing our proposed NTS-MTD technique because it can maximize MTTSF (even the number of attack paths toward decoy targets generated by DPNT is low) and minimize defense cost, while maintaining comparable service availability. 
\item Among the 12 schemes investigated for executing our proposed NTS-MTD technique, AS-DPNT/HS-DPNT (with AS/HS as the ``when-to-shuffle'' strategy and DPNT as the ``how-to-shuffle'' strategy) can achieve high MTTSF and deception effectiveness, while maintaining low defense cost and high service availability. Further, AS-DPNT/HS-DPNT are resilient against attackers with increasing intelligence capability of detecting decoy nodes. There exist an optimal setting for the system security vulnerability level threshold parameter and the maximum delay parameter for maximizing MTTSF. The analysis performed in this paper can help the system designer identify the best parameter setting under which MTTSF may be maximized.
\item Real node population has a high impact on the number of attack paths toward decoy targets, defense cost, and MTTSF. On the other hand, although decoy node population has a high impact on the number of attack paths toward decoy targets and defense cost, it does not largely improve MTTSF. 

\end{itemize}
As our {\bf future work}, we plan to explore the following research areas:
\begin{itemize}
\item We will develop distributed MTD operations with decentralized SDN controllers. To this end, we will develop criteria to divide an IoT network into multiple sub-networks which can be controlled by different SDN controllers with the aim of providing lightweight shuffling-based MTD solutions.
\item We will explore machine/deep learning-based approaches to compute an optimal network topology in network shuffling-based MTD. As an example, we may consider GNN to model complex relationships and learn information structured as graphs~\cite{Rusek2019SDN}. 
\item We will incorporate ML/DL-based network topology generation technology with the graphical security model (GSM) to determine the optimal network topology by reconstructing GSM and developing new security metrics for solution optimization (e.g., the average number of decoy nodes on an attack path). 
\end{itemize}

\bibliographystyle{IEEETran}
\bibliography{IoTsecurity}

\begin{thebibliography}{10}
\providecommand{\url}[1]{#1}
\csname url@samestyle\endcsname
\providecommand{\newblock}{\relax}
\providecommand{\bibinfo}[2]{#2}
\providecommand{\BIBentrySTDinterwordspacing}{\spaceskip=0pt\relax}
\providecommand{\BIBentryALTinterwordstretchfactor}{4}
\providecommand{\BIBentryALTinterwordspacing}{\spaceskip=\fontdimen2\font plus
\BIBentryALTinterwordstretchfactor\fontdimen3\font minus
  \fontdimen4\font\relax}
\providecommand{\BIBforeignlanguage}[2]{{%
\expandafter\ifx\csname l@#1\endcsname\relax
\typeout{** WARNING: IEEEtran.bst: No hyphenation pattern has been}%
\typeout{** loaded for the language `#1'. Using the pattern for}%
\typeout{** the default language instead.}%
\else
\language=\csname l@#1\endcsname
\fi
#2}}
\providecommand{\BIBdecl}{\relax}
\BIBdecl

\bibitem{Roman2013ComNet}
R.~Roman, J.~Zhou, and J.~Lopez, ``{On the features and challenges of security
  and privacy in distributed internet of things},'' \emph{Computer Networks},
  vol.~57, no.~10, pp. 2266--2279, 2013.

\bibitem{cho2018JDMS}
J.~H. Cho and N.~Ben-Asher, ``{Cyber defense in breadth: Modeling and analysis
  of integrated defense systems},'' \emph{The Journal of Defense Modeling and
  Simulation}, vol.~15, no.~2, pp. 147--160, 2018.

\bibitem{Ge2019Book}
M.~Ge, J.~H. Cho, B.~Ishfaq, and D.~S. Kim, \emph{Modeling and Design of Secure
  {Internet of Things}}.\hskip 1em plus 0.5em minus 0.4em\relax Wiley, 2020,
  ch. Modeling and Analysis of Proactive Defense Mechanisms for
  {Internet-of-Things}, iEEE Press.

\bibitem{Cho20-mtd-survey}
J.~{Cho}, D.~P. {Sharma}, H.~{Alavizadeh}, S.~{Yoon}, N.~{Ben-Asher}, T.~J.
  {Moore}, D.~S. {Kim}, H.~{Lim}, and F.~F. {Nelson}, ``Toward proactive,
  adaptive defense: {A} survey on moving target defense,'' \emph{IEEE
  Communications Surveys Tutorials}, pp. 1--1, 2020.

\bibitem{Hong2015TDSC}
J.~B. Hong and D.~S. Kim, ``{Assessing the Effectiveness of Moving Target
  Defenses using Security Models},'' \emph{IEEE Transactions on Dependable and
  Secure Computing}, vol.~13, no.~2, pp. 163--177, 2015.

\bibitem{Ge2017JNCA}
M.~Ge, J.~B. Hong, W.~Guttmann, and D.~S. Kim, ``{A framework for automating
  security analysis of the internet of things},'' \emph{Journal of Network and
  Computer Applications}, vol.~83, pp. 12--27, 2017.

\bibitem{Plaga2018ICST}
S.~{Plaga}, N.~{Wiedermann}, M.~{Niedermaier}, A.~{Giehl}, and T.~{Newe},
  ``{Future Proofing IoT Embedded Platforms for Cryptographic Primitives
  Support},'' in \emph{Proceedings of the 12th International Conference on
  Sensing Technology (ICST '18)}, 2018, pp. 52--57.

\bibitem{Casola2013AMM}
V.~Casola, A.~D. Benedictis, and M.~Albanese, ``{A Multi-Layer Moving Target
  Defense Approach for Protecting Resource-Constrained Distributed Devices},''
  in \emph{IRI}, 2013.

\bibitem{sherburne2014ACISRC}
M.~Sherburne, R.~Marchany, and J.~Tront, ``{Implementing moving target ipv6
  defense to secure 6lowpan in the internet of things and smart grid},'' in
  \emph{Proceedings of the 9th Annual Cyber and Information Security Research
  Conference (CISR '14)}.\hskip 1em plus 0.5em minus 0.4em\relax ACM, 2014, pp.
  37--40.

\bibitem{zeitz2017IOTDI}
K.~Zeitz, M.~Cantrell, R.~Marchany, and J.~Tront, ``{Designing a micro-moving
  target ipv6 defense for the internet of things},'' in \emph{Proceedings of
  the 2017 IEEE/ACM Second International Conference on Internet-of-Things
  Design and Implementation (IoTDI '17)}.\hskip 1em plus 0.5em minus
  0.4em\relax IEEE, 2017, pp. 179--184.

\bibitem{mahmood2016WFIOT}
K.~Mahmood and D.~M. Shila, ``{Moving target defense for Internet of Things
  using context aware code partitioning and code diversification},'' in
  \emph{Proceedings of the 2016 IEEE 3rd World Forum on Internet of Things
  (WF-IoT '16)}.\hskip 1em plus 0.5em minus 0.4em\relax IEEE, 2016, pp.
  329--330.

\bibitem{zeitz2018WCL}
K.~{Zeitz}, M.~{Cantrell}, R.~{Marchany}, and J.~{Tront}, ``{Changing the Game:
  A Micro Moving Target IPv6 Defense for the Internet of Things},'' \emph{IEEE
  Wireless Communications Letters}, vol.~7, no.~4, pp. 578--581, 2018.

\bibitem{Kouachi2018WINCOM}
A.~I. {Kouachi}, S.~{Sahraoui}, and A.~{Bachir}, ``{Per Packet Flow
  Anonymization in 6LoWPAN IoT Networks},'' in \emph{Proceedings of the 2018
  6th International Conference on Wireless Networks and Mobile Communications
  (WINCOM '18)}, 2018, pp. 1--7.

\bibitem{Nizzi2019IoT}
F.~{Nizzi}, T.~{Pecorella}, F.~{Esposito}, L.~{Pierucci}, and R.~{Fantacci},
  ``{IoT} security via address shuffling: The easy way,'' \emph{IEEE Internet
  of Things Journal}, vol.~6, no.~2, pp. 3764--3774, 2019.

\bibitem{Kahla2018TrustCom}
M.~{Kahla}, M.~{Azab}, and A.~{Mansour}, ``{Secure, Resilient, and
  Self-Configuring Fog Architecture for Untrustworthy IoT Environments},'' in
  \emph{Proceedings of the 2018 17th IEEE International Conference On Trust,
  Security And Privacy In Computing And Communications/ 12th IEEE International
  Conference On Big Data Science And Engineering (TrustCom/BigDataSE '18)},
  2018, pp. 49--54.

\bibitem{Wang2019IoT}
S.~{Wang}, H.~{Shi}, Q.~{Hu}, B.~{Lin}, and X.~{Cheng}, ``{Moving Target
  Defense for Internet of Things Based on the Zero-Determinant Theory},''
  \emph{IEEE Internet of Things Journal}, pp. 1--1, 2019.

\bibitem{Almohaimeed2019LISAT}
A.~{Almohaimeed}, S.~{Gampa}, and G.~{Singh}, ``{Privacy-Preserving IoT
  Devices},'' in \emph{Proceedings of the 2019 IEEE Long Island Systems,
  Applications and Technology Conference (LISAT '19)}, 2019, pp. 1--5.

\bibitem{Lin2019ICC}
G.~{Lin}, M.~{Dong}, K.~{Ota}, J.~{Li}, W.~{Yang}, and J.~{Wu}, ``{Security
  Function Virtualization Based Moving Target Defense of SDN-Enabled Smart
  Grid},'' in \emph{Proceedings of the 2019 IEEE International Conference on
  Communications (ICC '19)}, 2019, pp. 1--6.

\bibitem{Hamada2018IEMCON}
A.~O. {Hamada}, M.~{Azab}, and A.~{Mokhtar}, ``{Honeypot-like Moving-target
  Defense for secure IoT Operation},'' in \emph{Proceedings of the 2018 IEEE
  9th Annual Information Technology, Electronics and Mobile Communication
  Conference (IEMCON '18)}, 2018, pp. 971--977.

\bibitem{Vuppala2019GIoTS}
S.~{Vuppala}, A.~E. {Mady}, and A.~{Kuenzi}, ``{Rekeying-based Moving Target
  Defence Mechanism for Side-Channel Attacks},'' in \emph{Proceedings of the
  2019 Global IoT Summit (GIoTS '19)}, 2019, pp. 1--5.

\bibitem{Miyazaki2014ICNC}
T.~Miyazaki, S.~Yamaguchi, K.~Kobayashi, J.~Kitamichi, S.~Guo, T.~Tsukahara,
  and T.~Hayashi, ``{A Software Defined Wireless Sensor Network},'' in
  \emph{Proceedings of the IEEE 2014 International Conference on Computing,
  Networking and Communications (ICNC '14)}, 2014, pp. 847--852.

\bibitem{La2016IoTJ}
Q.~D. La, T.~Q.~S. Quek, J.~Lee, S.~Jin, and H.~Zhu, ``{Deceptive Attack and
  Defense Game in Honeypot-Enabled Networks for the Internet of Things},''
  \emph{IEEE Internet of Things Journal}, vol.~3, no.~6, pp. 1025--1035, 2016.

\bibitem{Anirudh2017ICCCSP}
M.~Anirudh, S.~A. Thileeban, and D.~J. Nallathambi, ``{Use of honeypots for
  mitigating DoS attacks targeted on IoT networks},'' in \emph{Proceedings of
  the 2017 International Conference on Computer, Communication and Signal
  Processing (ICCCSP '17)}.\hskip 1em plus 0.5em minus 0.4em\relax IEEE, 2017,
  pp. 1--4.

\bibitem{Dowling2017ISSC}
S.~Dowling, M.~Schukat, and H.~Melvin, ``A {ZigBee} honeypot to assess {IoT}
  cyberattack behaviour,'' in \emph{Proceedings of the 2017 28th Irish Signals
  and Systems Conference (ISSC '17)}.\hskip 1em plus 0.5em minus 0.4em\relax
  IEEE, 2017, pp. 1--6.

\bibitem{DeceptionVendors}
\BIBentryALTinterwordspacing
L.~Pingree. (2016) {Emerging Technology Analysis: Deception Techniques and
  Technologies Create Security Technology Business Opportunities}. [Online].
  Available:
  \url{https://www.gartner.com/doc/reprints?id=1-2LSQOX3\&ct=150824\&st=sb\&aliId=87768}
\BIBentrySTDinterwordspacing

\bibitem{Sheyner2002SP}
O.~Sheyner, J.~Haines, S.~Jha, R.~Lippmann, and J.~M. Wing, ``{Automated
  Generation and Analysis of Attack Graphs},'' in \emph{Proceedings of the 2002
  IEEE Symposium on Security and Privacy (SP '02)}.\hskip 1em plus 0.5em minus
  0.4em\relax IEEE Computer Society, 2002, pp. 273--284.

\bibitem{Saini2008JCSC}
V.~Saini, Q.~Duan, and V.~Paruchuri, ``{Threat Modeling using Attack Trees},''
  \emph{Journal of Computer Science in Colleges}, vol.~23, no.~4, pp. 124--131,
  2008.

\bibitem{Abie2012BAN}
H.~Abie and I.~Balasingham, ``{Risk-based Adaptive Security for Smart IoT in
  eHealth},'' in \emph{Proceedings of the 7th International Conference on Body
  Area Networks (BodyNets '12)}.\hskip 1em plus 0.5em minus 0.4em\relax ICST,
  2012, pp. 269--275.

\bibitem{Rullo2017ICDCS}
A.~Rullo, E.~Serra, E.~Bertino, and J.~Lobo, ``{Shortfall-Based Optimal
  Security Provisioning for Internet of Things},'' in \emph{Proceedings of 2017
  IEEE 37th International Conference on Distributed Computing Systems (ICDCS
  '17)}.\hskip 1em plus 0.5em minus 0.4em\relax IEEE, 2017, pp. 2585--2586.

\bibitem{Savola2012BAN}
R.~M. Savola, H.~Abie, and M.~Sihvonen, ``{Towards Metrics-driven Adaptive
  Security Management in e-Health IoT Applications},'' in \emph{Proceedings of
  the 7th International Conference on Body Area Networks (BodyNets '12)}.\hskip
  1em plus 0.5em minus 0.4em\relax ICST, 2012, pp. 276--281.

\bibitem{openflow2012ONF}
O.~N. Foundation, ``{OpenFlow Switch Specification (Version 1.3.0)},'' Tech.
  Rep., 2012.

\bibitem{de2015LAT}
B.~T. De~Oliveira, L.~B. Gabriel, and C.~B. Margi, ``{TinySDN: Enabling
  multiple controllers for software-defined wireless sensor networks},''
  \emph{IEEE Latin America Transactions}, vol.~13, no.~11, pp. 3690--3696,
  2015.

\bibitem{Galluccio2015INFOCOM}
L.~Galluccio, S.~Milardo, G.~Morabito, and S.~Palazzo, ``{SDN-WISE}: Design,
  prototyping and experimentation of a stateful {SDN} solution for {WIreless
  SEnsor} networks,'' in \emph{Proceedings of the 2015 IEEE Conference on
  Computer Communications (INFOCOM '15)}, 2015, pp. 513--521.

\bibitem{Lei2014VITAE}
T.~Lei, Z.~Lu, X.~Wen, X.~Zhao, and L.~Wang, ``{SWAN}: An {SDN} based campus
  {WLAN} framework,'' in \emph{Proceedings of the 2014 4th International
  Conference on Wireless Communications, Vehicular Technology, Information
  Theory and Aerospace Electronic Systems (VITAE '14)}, 2014, pp. 1--5.

\bibitem{Bernardos2014WirelessComm}
C.~J. Bernardos, A.~de~la Oliva, P.~Serrano, A.~Banchs, L.~M. Contreras,
  H.~Jin, and J.~C. Zuniga, ``{An Architecture for Software Defined Wireless
  Networking},'' \emph{IEEE Wireless Communications}, vol.~21, no.~3, pp.
  52--61, 2014.

\bibitem{Liu2015CommMag}
J.~Liu, Y.~Li, M.~Chen, W.~Dong, and D.~Jin, ``{Software-Defined Internet of
  Things for Smart Urban Sensing},'' \emph{IEEE Communications Magazine},
  vol.~53, no.~9, pp. 55--63, 2015.

\bibitem{ge2018FGCS}
M.~Ge, J.~B. Hong, S.~E. Yusuf, and D.~S. Kim, ``{Proactive defense mechanisms
  for the software-defined Internet of Things with non-patchable
  vulnerabilities},'' \emph{Future Generation Computer Systems}, vol.~78, pp.
  568--582, 2018.

\bibitem{gartner2003byzantine}
F.~C. G{\"a}rtner, ``{Byzantine failures and security: Arbitrary is not
  (always) random},'' Tech. Rep., 2003.

\bibitem{nist2005NVD}
\BIBentryALTinterwordspacing
NIST. (2005) {National Vulnerability Database (NVD)}. [Online]. Available:
  \url{https://nvd.nist.gov/}
\BIBentrySTDinterwordspacing

\bibitem{cho2017CST}
J.~H. Cho, Y.~Wang, R.~Chen, K.~S. Chan, and A.~Swami, ``{A survey on modeling
  and optimizing multi-objective systems},'' \emph{IEEE Communications Surveys
  \& Tutorials}, vol.~19, no.~3, pp. 1867--1901, 2017.

\bibitem{Cho17}
J.~H. Cho, Y.~Wang, I.~R. Chen, K.~S. Chan, and A.~Swami, ``{A Survey on
  Modeling and Optimizing Multi-Objective Systems},'' \emph{IEEE Communications
  Surveys Tutorials}, vol.~19, no.~3, pp. 1867--1901, 2017.

\bibitem{Rusek2019SDN}
K.~Rusek, J.~Su\'{a}rez-Varela, A.~Mestres, P.~Barlet-Ros, and
  A.~Cabellos-Aparicio, ``{Unveiling the Potential of Graph Neural Networks for
  Network Modeling and Optimization in SDN},'' in \emph{Proceedings of the 2019
  ACM Symposium on SDN Research (SOSR '19)}.\hskip 1em plus 0.5em minus
  0.4em\relax Association for Computing Machinery, 2019, pp. 140--151.

\end{thebibliography}

\appendix

\section{Decoy Path-based Optimization Algorithm}
In this appendix, we provide notations used in the optimization algorithm in the following and then present the algorithm in Algorithm~\ref{algo_opt}. 

\begin{itemize}
\item $m_i$: A node in the network for $i \in \{1, ..., n\}$ where $n$ is the total number of nodes
\item $M$: A set of nodes in the network, denoted by $M = \{m_1, ..., m_n\}$
\item $M_d$: A set of decoy nodes in the network, denoted by $M_d = \{m_{d_1}, \ldots, m_{d_n}\}$
\item $M_r$: A set of real nodes in the network, denoted by $M_r = \{m_{r_1}, \ldots, m_{r_n}\}$
\item $M_t$: A set of nodes considered as targets, denoted by $M_t \subset M$
\item $M_{dt}$: A set of decoy nodes considered as decoy targets, denoted by $M_{dt} \subset M_t$
\item $e_{m_i, m_j}$: An edge (connection) from $m_i$ to $m_j$
\item $m_i.con$: A list of out-degree connections of $m_i$ 
\item $\mathrm{getDecoys}(M, \zeta)$: A function that randomly selects a certain number of decoy IoT nodes from all decoy IoT nodes based on a probability $\zeta$ where each real IoT node is connected to at least half of the decoy nodes; a returned list is stored in $M_d$.
\item $\mathrm{checkCon}(m_i, m_j)$: Return true for $e_{m_i, m_j} > 0$; return false otherwise
\item $c$: A cost associated with shuffling edges which is incremented by 1 upon any edge changed (i.e., removal or addition)
\item $DP_{m_i}$: A set of paths toward decoy targets where node $m_i$ is an entry point
\item $\mathrm{traverseNet}(M)$: A function that takes a set of nodes in the network and returns $Dic_M$ with node $m_i$ as the key and $|DP_{m_i}|$ as the item where $m_i$ is a real IoT node
\item $\mathrm{Max}(Dic_M)$: A function that returns a list of nodes with maximum $|DP_{m_i}|$ 
\item $\mathrm{Min}(Dic_M)$: A function that returns a list of nodes with minimum $|DP_{m_i}|$ 
\end{itemize}

\begin{algorithm}[!ht]
\caption{Decoy Path-based Optimization Algorithm}
\label{algo_opt}
\begin{algorithmic}[1]
\Procedure{Decoy-Path-Optimization}{}
\For{$m_i \in M$}
    \If{$m_{i} \in M_r \wedge m_i \notin M_t$}
        \State{$M_d$ = $\mathrm{getDecoys}(M, \zeta)$}
        \For{$m_j \in M$}
            \If{$m_{j} \in M_d \wedge \mathrm{checkCon}(m_i, m_j)== \mathrm{false}$}
                \State{$e_{m_i, m_j}=1$} \Comment{Add an edge}
                \State{$c = c + 1$}
            \EndIf
        \EndFor
        \For{$m_k \in m_i.con$}
            \If{$m_k \in M_d \wedge m_k \notin M_{dt}$}
            \State{$e_{m_i, m_k}=0$} \Comment{Remove an edge}
            \State{$c = c + 1$}
            \EndIf 
        \EndFor 
    \EndIf 
\EndFor 
\State $Dic_M$ = $\mathrm{traverseNet}(M)$
\For{$m_i \in M$}
    \If{$m_i \in \mathrm{Min}(Dic_M)$}
        \For{$m_j \in M$}
            \If{$m_j \in \mathrm{Max}(Dic_M) \wedge m_j \notin m_i.con$}
                \State{$e_{m_i, m_j}=1$} \Comment{Add an edge}
                \State{$c=c+1$}            \EndIf 
        \EndFor 
    \EndIf 
\EndFor 
\EndProcedure
\end{algorithmic}
\end{algorithm}

In Algorithm~\ref{algo_opt}, we explain what each line does as follows:
\begin{itemize} 
\item[] {\bf line 2}: Go through each node in the network.
\item[] {\bf line 3}: Check two conditions: (i) whether node $m_i$ is a real node or not; and (ii) the node is a target or not.
\item[] {\bf line 4}: If node $m_i$ is a real node and not a target (i.e., a real IoT node), we randomly select a certain number of decoy IoT nodes and store them in $M_d$.
\item[] {\bf line 5}: Go through each node in the network.
\item[] {\bf line 6}: Check two conditions: (i) whether node $m_j$ belongs to $M_d$ or not; and (ii) whether the node is connected with $m_i$ or not.
\item[] {\bf lines 7-8}: Add an edge (connection) from $m_i$ to $m_j$ and increment the cost by 1.
\item[] {\bf line 11}: Go through each node in $mi.con$.
\item[] {\bf line 12}: Check two conditions: (i) whether node $m_k$ belongs to $M_d$; and (ii) whether the node belongs to $M_dt$ or not.
\item[] {\bf lines 13-14}: Remove the existing connection from $m_i$ to $m_k$ and increase the cost by 1.
\item[] {\bf line 19}: In the second step, we first compute a dictionary with each real IoT node as the key and number of decoy paths as the item. 
\item[] {\bf line 20}: Go through each node in the network.
\item[] {\bf line 21}: Check whether node $m_i$ has minimum number of decoy paths.
\item[] {\bf line 22}: Go through each node in the network.
\item[] {\bf line 23}: Check two conditions: (i) whether node $m_j$ has the maximum number of decoy paths or not; and (ii) the node is connected with $m_i$ or not.
\item[] {\bf lines 24-25}: Add an edge (connection) from $m_i$ to $m_j$ and increment the cost by 1.
\end{itemize}

\end{document}